\documentclass[12pt]{article}
\pdfoutput=1
\usepackage{latexsym, graphicx} 
\usepackage{amsmath}
\usepackage{amssymb}
\usepackage{amsfonts}
\usepackage{caption}
\usepackage{subcaption}
\usepackage{tensor}
\usepackage{cite}
\usepackage[utf8]{inputenc}
\allowdisplaybreaks

\usepackage[colorlinks=true, linkcolor=blue, bookmarks=true]{hyperref}
\newcommand{\arXiv}[1]{\href{http://www.arXiv.org/abs/#1}{#1}}

\makeatletter
\renewcommand\section{\@startsection {section}{1}{\z@}%
	{-3.5ex \@plus -1ex \@minus -.2ex}%nn
	{2.3ex \@plus.2ex}%
	{\normalfont\large\bfseries}}
\renewcommand\subsection{\@startsection{subsection}{2}{\z@}%
	{-3.25ex\@plus -1ex \@minus -.2ex}%
	{1.5ex \@plus .2ex}%
	{\normalfont\bfseries}}
\makeatother
\newcommand{\beq}{\begin{equation}}
\newcommand{\eeq}{\end{equation}}
\newcommand{\ber}{\begin{array}}
	\newcommand{\eer}{\end{array}}

\newcommand{\ena}{\end{eqnarray}}
\newcommand{\beqa}{\begin{eqnarray}}
\newcommand{\eeqa}{\end{eqnarray}}
\newcommand{\bea}{\begin{eqnarray}}
\newcommand{\eea}{\end{eqnarray}}

\newcommand{\be}{\begin{equation}}
\newcommand{\ee}{\end{equation}}

\begin{document}

\begin{titlepage}
	%\begin{flushright}
	%arXiv:xxxx.xxxx
	%\end{flushright}
	\vfill
	\begin{center}
		{\LARGE \bf Matrix Thermalization}
		
		\vskip 20mm
		
		{\large Ben Craps$\,^{a}$, Oleg Evnin$\,^{b,a}$, K\'evin Nguyen$\,^{a}$}
		
		\vskip 15mm
		
		$^a$ Theoretische Natuurkunde, Vrije Universiteit Brussel (VUB), and \\ International Solvay Institutes, Pleinlaan 2, B-1050 Brussels, Belgium \\
		\vskip 3mm
		$^b$ Department of Physics, Faculty of Science, Chulalongkorn University,\\
			Thanon Phayathai, Pathumwan, Bangkok 10330, Thailand\\
		
		\vskip 15mm
		
		{\small\noindent  {\tt Ben.Craps@vub.ac.be, oleg.evnin@gmail.com, Kevin.Huy.D.Nguyen@vub.ac.be}}
		
	\end{center}
	\vfill

	\begin{center}
		{\bf ABSTRACT}
		\vspace{3mm}
	\end{center}
	
		Matrix quantum mechanics offers an attractive environment for discussing gravitational holography, in which both sides of the holographic duality are well-defined. Similarly to higher-dimensional implementations of holography, collapsing shell solutions in the gravitational bulk correspond in this setting to thermalization processes in the dual quantum mechanical theory. We construct an explicit, fully nonlinear supergravity solution describing a generic collapsing dilaton shell, specify the holographic renormalization prescriptions necessary for computing the relevant boundary observables, and apply them to evaluating thermalizing two-point correlation functions in the dual matrix theory.

\vfill
	
\end{titlepage}	
	
\tableofcontents	

%\newpage	
%%%%%%%%%%%%%%%%%%%%%%%%%%%%%%%%%%%%%%%%%%%%%%%%%%%%%%%
\section{Introduction}

In recent years, gauge/gravity duality, also known as ``holography,'' has emerged as a rare tool for the study of strongly coupled systems far from equilibrium. Originally motivated by the creation of a quark gluon plasma in ultrarelativistic heavy ion collisions, many authors have used the AdS/CFT correspondence to study what happens when energy is suddenly injected in a strongly coupled quantum field theory. Interesting results include thermalization times short enough to be compatible with experiments \cite{Kovchegov:2007pq, Albacete:2008vs, Chesler:2008hg, Beuf:2009cx, AbajoArrastia:2010yt,Albash:2010mv,Balasubramanian:2010ce,Heller:2011ju,Heller:2012km, Balasubramanian:2013rva}, a thermalization pattern in which short-wavelength modes thermalize first \cite{Lin:2008rw, Balasubramanian:2010ce}, and new insights in the spreading of entanglement entropy after a 1+1d quantum quench \cite{AbajoArrastia:2010yt, Balasubramanian:2010ce, Balasubramanian:2011at, Liu:2013iza, Asplund:2015eha}.

It is interesting to ask whether holography can be used to make predictions for the thermalization of systems that can also be studied using conventional techniques. If so, this would provide a framework in which holography can be quantitatively tested in a far-from-equilibrium regime. With this question in mind, we will study holographic thermalization in BFSS matrix theory \cite{deWit:1988wri,Banks:1996vh}, a quantum-mechanical model of $N\times N$ matrices which, in the $N\to\infty$ limit, has been proposed as a nonperturbative definition of M-theory in asymptotically flat backgrounds \cite{Banks:1996vh}. Our considerations will revolve around the relation of this model \cite{Polchinski:1999br} to a non-conformal version of the AdS/CFT correspondence \cite{Itzhaki:1998dd, Sekino:1999av, Wiseman:2008qa, Kanitscheider:2008kd}. It has also appeared in recent discussions of ``fast scrambling'' \cite{Sekino:2008he}, the fast spreading of information that is added ``locally'' (e.g., in one matrix component). While our focus will be on far-from-equilibrium processes driven by energy injection, a simpler holographic setup involving the same matrix theory has been previously studied, in a way involving extensive numerical simulations, in a sequence of papers including \cite{Anagnostopoulos:2007fw,Catterall:2008yz,Hanada:2008ez,Catterall:2009xn,Kadoh:2015mka,Filev:2015hia}. In those considerations, a stationary black hole was introduced in the gravitational bulk, corresponding to thermodynamic equilibrium, rather than thermalization, in the dual matrix theory. Further analytic considerations of matrix theory thermodynamics can be found in \cite{Smilga:2008bt,Wiseman:2013cda}. In \cite{Morita:2013wfa,Morita:2014ypa}, dynamics of moduli fields has been explored as a tool to probe thermal properties of higher-dimensional super-Yang-Mills theories, though applications to the case of matrix theory are less straightforward.

Despite the apparent simplicity of the BFSS matrix theory, which involves quantum mechanics rather than quantum field theory, real-time evolution of this model in the appropriate strong coupling regime presently appears to be out of reach of conventional techniques, even for small values of $N$. The nine scalar matrices and their fermionic partners contain too many degrees of freedom to allow direct diagonalization, and the interactions between the various matrix elements appear to be too nonlocal for variational techniques such as tensor network methods to be directly applicable. It would be really nice if these or other techniques could be developed up to the point where they can capture matrix theory, first for small $N$ and later for larger values of $N$, in order to allow detailed comparison with holography. (We would like to mention an intriguing attempt to tackle the quantum dynamics of a simpler bosonic matrix theory undertaken in \cite{Hubener:2014pfa}.) In the meantime, numerical simulations have been carried out in another regime, where the matrix theory can be treated classically \cite{Asplund:2012tg, Aoki:2015uha, Gur-Ari:2015rcq}. (A similar study of the related BMN matrix model can be found in \cite{Asplund:2011qj}.) This is a simplification which would not arise in higher dimensions; see \cite{Gur-Ari:2015rcq} and references therein. One motivation is that, according to numerical simulations, there is no phase transition between the different regimes \cite{Anagnostopoulos:2007fw,Catterall:2008yz,Hanada:2008ez,Catterall:2009xn,Kadoh:2015mka,Filev:2015hia}, so some qualitative features can be expected to be similar \cite{Gur-Ari:2015rcq}. Further studies of semiclassical processes in the matrix theory revolving around the idea of continuity from weak to strong coupling can be found in \cite{Iizuka:2013yla,Iizuka:2013kha}.

In this paper, we use D0-brane holography \cite{Itzhaki:1998dd, Sekino:1999av, Wiseman:2008qa, Kanitscheider:2008kd} to study far-from-equilibrium evolution of matrix theory after a sudden injection of energy. In higher-dimensional AdS/CFT, a simple way to inject energy in a holographic field theory is by briefly turning on and off a homogeneous source, for instance for an anisotropic component of the stress tensor \cite{Chesler:2008hg}, for an electric current \cite{Horowitz:2013mia} or for a scalar operator \cite{Bhattacharyya:2009uu}. In the bulk, this corresponds to turning on nontrivial boundary conditions for the corresponding bulk field. For a small-amplitude scalar perturbation, an approximate AdS-Vaidya spacetime was found \cite{Bhattacharyya:2009uu}, and this has become an often-used toy model for homogeneous, isotropic energy injection. Interestingly, the electric field perturbation of \cite{Horowitz:2013mia} yields an exact AdS-Vaidya spacetime.

For D0-brane holography, if one restricts to an ansatz that is spherically symmetric in the ``internal'' directions (transverse to the D0-brane worldvolume), the supergravity field equations simplify to those of a dilaton-gravity model \cite{Wiseman:2008qa, Kanitscheider:2008kd} coupled to an additional scalar (the ``breathing mode" of the internal sphere) \cite{Wiseman:2008qa}. We will explicitly solve the dilaton-gravity equations (with the breathing mode set to zero) with an arbitrary boundary profile for the dilaton (corresponding to an arbitrary source for the dual operator). As a consequence of the lack of dynamical degrees of freedom in 2d dilaton-gravity, if one turns a source on and off, the late-time bulk metric agrees with the early-time bulk metric, and no net energy was injected. We will find, however, that one can inject energy in the system by considering a boundary condition for the dilaton that is constant at early times and evolves to a different constant value at late times. In field theory language, this corresponds to starting with a thermal state and ending with a thermal state at a different temperature (and with a different value of the coupling constant).

Concretely, in Appendix~\ref{App:Background Solution} we derive the following exact analytic solution of \textrm{II}A supergravity, expressed in a \textit{dual frame} in which the 2d metric is asymptotically AdS (see Section~\ref{section:Review}, where more details will be provided):
\begin{align}
\label{70}
ds^2_{dual}&=-\frac{1}{x^2}\left[2\ dvdx+\left(1+\frac{20}{21}\dot{\phi}^{(0)}(v)\ x-M_0\ e^{\frac{4}{3}\phi^{(0)}(v)}x^{14/5}\right)dv^2\right]+\frac{25}{4}d\Omega_8^2,\\
\label{71}
\phi(v,x)&=\phi^{(0)}(v)+\frac{21}{10}\log x,
\end{align}
where $M_0$ is a mass parameter and $\phi^{(0)}(v)$ is a function that one is free to choose as Dirichlet boundary condition for the dilaton field $\phi$ and which we also call dilaton \textit{source}. The metric \eqref{70} describes a black hole with mass $M=M_0\ e^{\frac{4}{3}\phi^{(0)}}$ and is asymptotically $AdS_2 \times S^8$  for $x\rightarrow 0$, where the timelike boundary of $AdS_2$ is located. Provided that we have $M_0\neq 0$, a non-constant dilaton boundary value $\phi^{(0)}(v)$ will effectively result in a non-constant mass term in the metric. 

Even though the above collapsing solution allows arbitrary energy injection patterns, in this paper, we will mostly consider the thin-shell limit in which a black hole spacetime of some initial temperature is glued to a black hole spacetime of higher temperature at a null surface $v=v_0$. This can be achieved by assuming the following profile for the dilaton source:
\begin{align}
\label{72a}
&v<v_0: \qquad \phi^{(0)}(v)=\phi_0,\\
& \nonumber \\
\label{73a}
&v>v_0: \qquad \phi^{(0)}(v)=0,
\end{align}
with $\phi_0$ a negative constant. For computational simplicity, we will often consider the $\phi_0\to-\infty$ limit in which the initial temperature vanishes and the early-time geometry is vacuum $AdS_2$: 
\begin{align}
\label{72}
&v<v_0: \qquad \begin{cases}
\phi^{(0)}(v)\rightarrow-\infty,\\
\dot{\phi}^{(0)}(v)=0,
\end{cases}\\
& \nonumber \\
\label{73}
&v>v_0: \qquad \phi^{(0)}(v)=0.
\end{align}

At least within an energy range to be discussed in the next section, this solution is holographically dual to matrix theory excited (``quenched'') away from equilibrium through energy injection. However, the solution \eqref{70}-\eqref{71} does not describe propagating degrees of freedom and will not be sufficient for computing non-trivial correlation functions. As a dynamical probe of this background, we then consider fluctuations in the size of the compact $S^8$, i.e., the breathing mode. This mode has already been considered in previous holographic works \cite{Sekino:1999av,Wiseman:2008qa} and has been identified in \cite{Sekino:1999av} to be dual to a matrix theory operator $T^{--}$, to be defined in \eqref{56}, by matching of \textit{generalized conformal scaling dimensions} \cite{Jevicki:1998ub,Jevicki:1998yr,Yoneya:1999bb}. Our setup will allow us to holographically compute its retarded two-point correlation function in the quenched dual state, thereby providing a first non-trivial observable which, in the future, one may hope to compare with direct matrix theory computations. Predictably, the late-time behavior of this correlation function is dominated by the lowest quasinormal mode of the final state black hole.

$AdS_2$ holography has a reputation for being very subtle and relatively poorly understood (see \cite{Strominger:1998yg, Maldacena:1998uz, Almheiri:2014cka, Jensen:2016pah, Maldacena:2016upp, Engelsoy:2016xyb, Grumiller:2016dbn, Cvetic:2016eiv} for a sampling of the literature, with an emphasis on recent discussions), so one might wonder why we did not run into problems when considering $AdS_2$ backgrounds and excitations thereof. To see the difference between our D0-brane holography and what is usually referred to as $AdS_2$ holography, note that our $AdS_2$ solution arises in the dual frame, in which the effective dilaton-gravity action takes the form \eqref{61} with constant $b=25/4$. This action has a nontrivial dilaton kinetic term, the removal of which would require a dilaton-dependent rescaling of the metric. After such a rescaling, the metric would no longer be asymptotically $AdS_2$. This should be contrasted with conventional $AdS_2$ holography, which considers asymptotically $AdS_2$ solutions in the frame without a dilaton kinetic term, which turns out to be more subtle. More specifically, subtleties such as the absence of finite-energy
excitations for fixed asymptotics arise for $AdS_2$ solutions with constant
dilaton.\footnote{We thank Ioannis Papadimitriou for a useful discussion on this point.} In the theory we are considering here, the dilaton field depends at least on the radial coordinate for all solutions of the equations of
motion, and holography works in a way similar to the running dilaton
solutions considered in \cite{Cvetic:2016eiv}.

The paper is organized as follows. In Section~\ref{section:Review} we review the duality between matrix theory and \textrm{II}A supergravity originally conjectured in \cite{Itzhaki:1998dd}, setting thereby the conventions that are going to be used throughout this work. In Section~\ref{section:Dimensional Reduction}, we study the bosonic part of type \textrm{II}A supergravity with asymptotically $AdS_2 \times S^{8}$ geometry in the dual frame. In order to simplify the problem as much as possible, we only consider spherically symmetric solutions, leading to a two-dimensional effective theory describing the metric, the dilaton and the breathing mode accounting for the $S^{8}$ size dynamics. This mode is the only propagating physical degree of freedom in that system, and will be our probe for computing non-trivial correlations functions in the quenched dual state. It is important to note that the breathing mode cannot be considered nonperturbatively as noted in \cite{Wiseman:2008qa}, because it deforms the boundary away from $AdS_2$ (see Appendix \ref{App:Asymptotics}). In terms of matrix theory, its dual operator $T^{--}$ is irrelevant and cannot be sourced nonperturbatively. Nonetheless, a proper perturbative treatment of the breathing mode is expected to correctly reproduce the correlation functions of the dual matrix theory operator \cite{Wiseman:2008qa,vanRees:2011fr}. In Section \ref{section:Holographic Renormalization} we perform the holographic renormalization procedure \cite{Skenderis:2002wp}. Knowing the fields' asymptotics near the $AdS_2$ boundary as well as the on-shell effective action, local boundary counterterms are added in order to cancel boundary divergences. This is part of the precise holographic description of matrix theory. In earlier works holographic renormalization has been performed for various cases, including non-linear gravity-dilaton solutions \cite{Kanitscheider:2008kd} and breathing mode perturbations around pure $AdS_2 \times S^{8}$ \cite{Wiseman:2008qa}. Related work also includes \cite{Ortiz:2014aja}. A general discussion of holographic renormalization in the presence of irrelevant operators deforming the AdS boundary can be found in \cite{vanRees:2011fr}. Here we consider breathing mode perturbations around non-linear gravity-dilaton background solutions, allowing in particular for time-dependent backgrounds of the form \eqref{70}-\eqref{71}. 

In Section~\ref{section:Propagators} we compute the retarded boundary-to-bulk propagator of the breathing mode in the case of pure $AdS_2$ and in the more interesting case of the thin-shell solution \eqref{70}-\eqref{73}, which is dual to a quenched state in matrix theory. For the retarded propagator in the latter case, we use numerical evolution and show that its asymptotic value near the boundary is rapidly dominated by a single decaying and oscillating mode after crossing of the shell located at $v=v_0$. We also show that the associated single complex frequency dominating the retarded two-point function in this quenched state with final temperature $T$ coincides with the lowest quasinormal mode frequency of breathing mode fluctuations around a static black hole at the same temperature $T$. The first and second quasinormal mode frequencies are therefore computed in Appendix \ref{App:LowestQNM} using a numerical shooting method. Using these results, we holographically derive in Section~\ref{section:Linear Response} the retarded non-equal-time two-point function of $T^{--}$. Earlier computations of holographic two-point functions in equilibrium states (as opposed to our far-from-equilibrium setting) can be found in \cite{Sekino:1999av,Matsuo:2013jda,Sekino:2000mg}.

%%%%%%%%%%%%%%%%%%%%%%%%%%%%%%%%%%%%%%%%%%%%%%%%%%%%%%%%%%
\section{Review of \textrm{II}A Supergravity - Matrix Theory Duality} \label{section:Review}

We start by reviewing the duality originally presented in \cite{Itzhaki:1998dd}, looking only at terms relevant for the present work and setting our conventions. Useful references include \cite{Boonstra:1998mp,Skenderis:1999bs,Sekino:1999av}. The bosonic part of the 10d type \textrm{II}A supergravity action in string frame is
\begin{equation}
S_{string}=\frac{1}{(2 \pi)^7 g_s^2 \alpha'^4}\int d^{10}x\ \sqrt{-g} \left[e^{-2 \phi}(R+4(\partial \phi)^2)-\frac{1}{4}F^2\right],
\end{equation}
with $g_s$ and $\sqrt{\alpha'}=l_s$ being the string coupling and the string length, respectively. This action involves the metric, a scalar dilaton $\phi$ and a gauge potential $C_M$ with field strength $F_{MN}=\partial_M C_N - \partial_N C_M$ and density $F^2\equiv F_{MN}F^{MN}$. This system admits a solution representing $N$ coincident electric D-particles at the origin \cite{Horowitz:1991cd}:
\begin{align}
\label{57}
ds^2_{string}&=-H^{-1/2} dt^2+H^{1/2}dx_i dx^i,\\
\label{58}
e^{\phi}&=H^{3/4},\\
\label{59}
C_0&=H^{-1}-1,
\end{align} 
where $H$ is a single-centered harmonic function on the Euclidean space labeled by Cartesian coordinates $x^i$, given by
\begin{equation}
\label{64}
H=1+\frac{Q}{r^7}, \qquad r^2\equiv \sum_{i=1}^{9} x_i^2, \qquad Q=60 \pi^3 g_sN (\alpha')^{7/2}.
\end{equation}

It has been conjectured that the near-horizon limit or decoupling limit of the above D-particle background is dual to matrix theory \cite{Itzhaki:1998dd}. Explicitly, this decoupling limit is 
\begin{align}
\label{54}
g_s\rightarrow 0, \qquad \alpha' \rightarrow 0, \qquad U\equiv\frac{r}{\alpha'}=\text{fixed}, \qquad g_{YM}^2N=\text{fixed},
\end{align}
where the energy is kept fixed while taking the limit, and the Yang-Mills coupling of matrix theory is identified with
\begin{equation}
g_{YM}^2=4\pi^2g_s (\alpha')^{-3/2}.
\label{gYMdef}
\end{equation}

Performing a Weyl transformation on the string frame metric \eqref{57} while defining $\beta_0\equiv \frac{4}{25}(15\pi)^{2/7}$, one can go to the so-called \textit{dual frame} \cite{Boonstra:1998mp} 
\begin{align}
\label{55}
ds^2_{dual}&\equiv \beta_0^{-1}\ \alpha'^{-10/7}\left(g_{YM}^2 N e^{\phi}\right)^{-2/7}ds^2_{string}\\
\label{8}
&=\frac{25}{4}\left[-\frac{U^5}{15\pi g_{YM}^2N} dt^2+U^{-2}dU^2+d\Omega^2_8\right],
\end{align}
in which the action reads
\begin{align}
S_{dual}=\frac{\beta_0^4\ (g_{YM}^2N)^{8/7}}{(2 \pi)^3 (\alpha')^{9/7} g_{YM}^4}\int d^{10}x\ \sqrt{-g} \left[e^{-\frac{6}{7} \phi}(R+\frac{16}{49} (\partial \phi)^2)-\frac{e^{\frac{6}{7} \phi}}{4 \beta_0 (\alpha')^{10/7} (g_{YM}^2 N)^{2/7}}F^2\right].
\end{align}
For further simplification, we apply the following fields redefinitions, 
\begin{align}
e^{\tilde{\phi}}&\equiv\beta_1^{-1}\ (\alpha')^{3/2}(g_{YM}^2N)^{3/10}\ e^{\phi},\\
\tilde{C_0}&\equiv\beta_0^{-1/2}\beta_1^{\frac{6}{7}}\ (\alpha')^{-2}(g_{YM}^2N)^{-2/5}\ C_0,
\end{align}
where we define $\beta_1\equiv\frac{5\times 5^{4/5}}{4\times2^{1/10}(3\pi)^{3/10}}$, bringing the action to the form
\begin{align}
\label{9}
S_{dual}=\frac{\beta_0^4\ \beta_1^{-\frac{6}{7}}(g_{YM}^2N)^{7/5}}{(2 \pi)^3 g_{YM}^4}\int d^{10}x\ \sqrt{-g} \left[e^{-\frac{6}{7} \tilde{\phi}}(R+\frac{16}{49} (\partial \tilde{\phi})^2)-\frac{1}{4}e^{\frac{6}{7} \tilde{\phi}}\tilde{F}^2\right].
\end{align}
\newline
Finally, by performing the coordinate redefinition
\begin{equation}
\label{50}
z^2=\frac{12\pi}{5} g_{YM}^2NU^{-5},
\end{equation}
the D-particle background in the decoupling limit \eqref{8} becomes
\begin{align}
\label{51}
ds^2_{dual}&=\frac{1}{z^2}\left(-dt^2+dz^2\right)+\frac{25}{4}d\Omega^2_8,\\
\label{52}
e^{\tilde{\phi}}&=z^{21/10},\\
\label{53}
\tilde{F}_{0z}&=\frac{14}{5}z^{-19/5} \quad \Rightarrow \quad  \tilde{F}^2=-2\left(\frac{14}{5}\right)^2 z^{-18/5}.
\end{align}
The geometry of this dual frame metric is therefore manifestly that of $AdS_2 \times S^8$, with asymptotic boundary located at $z=0$. Moreover, it has been argued in \cite{Peet:1998wn} that the coordinate $z$ of the dual frame is proportional to the inverse energy scale of the boundary theory, making the dual frame a natural \textit{holographic frame}. It is also important to note that all relevant expressions appearing in \eqref{9}-\eqref{53} depend on quantities held fixed in the decoupling limit \eqref{54}.

The validity of the supergravity description requires the string frame curvature and dilaton to be small \cite{Itzhaki:1998dd}:
\begin{align}
\alpha' R_{string} &\sim \sqrt{\frac{U^{3}}{g_{YM}^2N}} \sim \left(z^3 g_{YM}^2N\right)^{-1/5} \ll 1,\\
g_s\ e^{\phi}&\sim \frac{1}{N} \left(z^3 g_{YM}^2 N\right)^{7/10}\ll 1.
\end{align}
These conditions can be conveniently summarized by
\begin{equation}
1\ll z^3 g_{YM}^2N\ll N^{10/7}.
\end{equation}
We see that near the AdS boundary (in the UV regime of the dual field theory), supergravity loses its validity because of strong curvature; in fact, very near the boundary perturbative matrix theory becomes valid. On the other hand, string loop corrections invalidate classical supergravity for large values of $z$ (in the IR regime of the dual field theory). The regime of validity of supergravity can be made parametrically large, however, by working in the large $N$ limit and at large values of the 't Hooft coupling $g_{YM}^2N$. There, the gravitational description can be trusted for matrix theory observables that are neither dominated by the extreme UV or IR parts of its spectrum. Indeed, in \cite{Hanada:2011fq} Monte Carlo computations of various two-point functions in the matrix theory ground state have been shown to agree well with earlier holographic predictions, at least when neither of the two extreme regimes mentioned above are considered.

On the other side of the duality, matrix theory is the non-abelian $U(N)$ gauge theory describing the decoupled low-energy dynamics of $N$ probe D-particles on a flat background. Its action can be written as \cite{Taylor:1999gq}
\begin{equation}
S_{BFFS}=\frac{1}{2g_{YM}^2}\int dt\ \text{Tr}\left[D_tX_iD_tX^i+\frac{1}{2}\left[X_i,X_j\right]^2+\theta^T \left(iD_t-\gamma_i\left[X^i,\ \cdot \ \right]\right)\theta\right],
\end{equation}  
where $X_i$ with $i=1,...,9$ are $N \times N$ hermitian matrices, $\theta$ is a 16-dimensional $SO(9)$ spinor whose components are $N \times N$ Grassmann matrices, $\gamma_i$ are the associated $SO(9)$ gamma matrices and $D_t\equiv \partial_t - i \left[A ,\ \cdot \ \right]$ is the covariant derivative associated to the $U(N)$ gauge field $A$.

Although in the present discussion we considered a supergravity background with pure $AdS_2\times S^{8}$ geometry, the duality applies in principle to any background with \textit{asymptotically} $AdS_2\times S^{8}$ geometry (in the dual frame). The black D0-brane solution giving rise in the decoupling limit to an $AdS_2$ black hole has for example been considered in the original paper of Itzhaki et \textit{al} \cite{Itzhaki:1998dd}, and associated holographic computations have been done in \cite{Matsuo:2013jda}.

%%%%%%%%%%%%%%%%%%%%%%%%%%%%%%%%%%%%%%%%%%%%%%%%%%%%%%%%%
\section{Dimensional Reduction of \textrm{II}A Supergravity}\label{section:Dimensional Reduction}

We are now going to study \textrm{II}A supergravity for asymptotically $AdS_2 \times S^{8}$ geometry in the dual frame, and we will restrict our attention to solutions that are spherically symmetric ($l=0$ Kaluza-Klein modes). For this, we use the dual frame action \eqref{9}, dropping for simplicity the tilde from all symbols together with the fixed overall factor, which can be easily restored:
\begin{align}
\label{60}
&S_{dual}=\int d^{10}x\ \sqrt{-g} \left[e^{-\frac{6}{7} \phi}(R+\frac{16}{49} (\partial \phi)^2)-\frac{1}{4}e^{\frac{6}{7} \phi}F^2\right] + S_{GH},\\
&S_{GH}=2\int d^{9}x\ \sqrt{-\gamma}\ e^{-\frac{6}{7}\phi}K,
\end{align}
where $\gamma$ is the induced metric at the asymptotic boundary whose extrinsic curvature is $K$, $S_{GH}$ being the standard Gibbons-Hawking term necessary to have a well-posed variational principle with Dirichlet boundary conditions on variations of the metric.

The equations of motion obtained from varying this action together with the Bianchi identity associated to the gauge field read 
\begin{multline}
\label{3}
-R_{MN}+\frac{20}{49}\partial_M\phi\partial_N\phi-\frac{6}{7}\nabla_M\partial_N\phi
+\frac{1}{2}g_{MN}\left[R-\frac{8}{7}(\partial\phi)^2+\frac{12}{7} \nabla^2\phi\right]\\
-\frac{1}{8}e^{\frac{12}{7}\phi}F^2g_{MN}+\frac{1}{2}e^{\frac{12}{7}\phi}g^{M'N'}F_{MM'}F_{NN'}=0,
\end{multline}
\begin{align}
\label{4}
R-\frac{16}{49} (\partial \phi)^2+\frac{16}{21} \nabla^2\phi+\frac{1}{4}e^{\frac{12}{7}\phi}F^2=0,
\end{align}
\begin{align}
\label{1}
\nabla_M\left(e^{\frac{6}{7}\phi}F^{MN}\right)=0,\\
\label{2}
\nabla_{[P} F_{MN]}=0.
\end{align}

We now postulate that the 10-dimensional spacetime is the warped product of a 2-dimensional Lorentzian spacetime and a Euclidean $S^8$, with the following block diagonal metric form:
\begin{equation}
\label{62}
ds^2_{dual}(x^\rho,x^i)\equiv g_{MN}dx^M dx^N=g_{\mu\nu}(x^\rho) dx^{\mu} dx^{\nu} + b(x^\rho)d\Omega_8^2(x^i).
\end{equation}
We use Greek indices for labeling the two-dimensional spacetime and Latin indices for the $S^8$ part. This metric ansatz almost decouples the two subspace metrics, except for the breathing mode $b(x^\rho)$. This mode has been identified in \cite{Sekino:1999av} as being dual to the matrix theory operator \cite{Kabat:1997sa}
\begin{align}
\label{56}
T^{--}&= \frac{1}{96\ g_{YM}^2}\ \text{STr}\left[\mathcal{F}\right],
\end{align}
where we define 
\begin{align}
\mathcal{F}&\equiv 24F_{0i}F_{0i}F_{0j}F_{0j}+24F_{0i}F_{0i}F_{jk}F_{jk}+96F_{0i}F_{0j}F_{ik}F_{kj}\\
&\hspace{0.5cm} +24F_{ij}F_{jk}F_{kl}F_{li}-6F_{ij}F_{ij}F_{kl}F_{kl},\nonumber \\
F_{0i}&\equiv D_t X_i,\\
F_{ij}&\equiv i \left[X_i,X_j\right].
\end{align}
In (\ref{56}), $\text{STr}$ stands for the symmetrized trace operation, consisting of full symmetrization over all possible orderings of the $F$-operator factors, followed by taking the ordinary trace. This identification has been performed by matching of generalized conformal scaling dimensions \cite{Jevicki:1998ub,Jevicki:1998yr,Yoneya:1999bb}. It is important to keep in mind that, in the conventions we are using, $g_{YM}$ in (\ref{56}) is given by the constant value (\ref{gYMdef}), which does not include a dilaton dependence, and hence does not depend on time.

We also choose an ansatz for the gauge field energy-momentum tensor $F$, often referred to as a Freund-Rubin ansatz \cite{Freund:1980xh}:
\begin{equation}
\label{48}
F_{MN}=\begin{cases}
A(x^R)\ \sqrt{-g(x^{\rho})}\ \epsilon_{\mu\nu} \quad &M=\mu,N=\nu,\\
0 \quad &\text{otherwise},
\end{cases}
\end{equation}
for which $A(x^R)$ is a priori a function of all coordinates, and $\epsilon_{\mu\nu}$ is the Levi-Civita symbol of the two-dimensional spacetime associated with coordinates $x^\rho$. This ansatz amounts to assuming a purely radial electric field. Furthermore we assume that the dilaton is independent of the $S^8$ coordinates,
\begin{equation}
\label{63}
\phi=\phi(x^\mu).
\end{equation}

The function $A$ appearing in  \eqref{48} is actually expressed by equations \eqref{1}-\eqref{2} through $b$ and $\phi$. In this situation, it is advantageous to develop an effective theory involving the two-dimensional metric $g_{\mu\nu}$, $b$ and $\phi$. While we give the details of this dimensional reduction in  Appendix \ref{App:Reduction}, we point out here a subtlety that arises when performing this dimensional reduction at the level of the action.

In an asymptotically AdS setting, $g_{\mu\nu}$, $b$ and $\phi$ satisfy Dirichlet boundary conditions at the conformal boundary of AdS. If \eqref{48} is imposed (with $A$ expressed through $b$ and $\phi$), this will imply that the gauge field strength $F$ (rather than the gauge potential $C$) satisfies Dirichlet boundary conditions. If we want to produce an effective action for $g_{\mu\nu}$, $b$ and $\phi$ we should thus start with a ten-dimensional action which allows a consistent variational principle based on variations of $F$ satisfying Dirichlet boundary conditions. This requires adding the following boundary term to \eqref{60}:
\begin{align}
\label{108}
S_{extra}&=\int d^9x\ \sqrt{-\gamma}\ n_{M} \left[e^{\frac{6}{7}\phi} F^{MN}C_{N}\right]\\
\label{109}
&=\int d^{10}x\ \sqrt{-g}\ \nabla_M \left[e^{\frac{6}{7} \phi}F^{MN}\right]C_N+\frac{1}{2}\int d^{10}x\ \sqrt{-g}\ e^{\frac{6}{7} \phi}F^2,
\end{align}
with $n^{M}$ being the outward-pointing normal unit vector of the boundary. Similar boundary terms have been considered in \cite{Duncan:1989ug,Groh:2012tf}. Equation \eqref{109} highlights the fact that the boundary term effectively corrects the coefficient of the Maxwell term when the action is evaluated on field configurations satisfying the gauge field equations of motion. This feature will correspondingly be reflected in the action of the dimensionally reduced theory, in which the gauge field is expressed through $g_{\mu\nu}$, $b$ and $\phi$. (One can straightforwardly verify that naively substituting the two-dimensional ansatz in the ten-dimensional action without taking into account the boundary term subtlety would have resulted in an action that no longer reproduces the correct equations of motion of the dimensionally reduced theory.)

Putting everything together (see Appendix \ref{App:Reduction} for the details) one finds that the action \eqref{60} is effectively reduced to two dimensions,
\begin{align}
\label{61}
S_{eff}&=\frac{2 \pi^{9/2}}{\Gamma \left(\frac{9}{2}\right)}\int d^{2}x\, \sqrt{-g}\, e^{-\frac{6}{7}\phi} \left[b^4R+56 b^{3}+14b^{2}(\partial b)^2-\frac{48}{7} b^{3}\partial_{\mu}b\, \partial^{\mu}\phi+\frac{16}{49} b^{4} (\partial \phi)^2-Cb^{-4}\right]\nonumber\\
&+\frac{4 \pi^{9/2}}{\Gamma \left(\frac{9}{2}\right)}\int d^{1}x\ \sqrt{-\gamma}\ e^{-\frac{6}{7}\phi}b^4 K,
\end{align}
where all geometric quantities are constructed from the two-dimensional metric, and $C=\frac{1}{2}(\frac{14}{5})^2 (\frac{25}{4})^8$. Furthermore one can check that this effective action leads to the correct equations of motion. We are effectively considering two scalars $\phi$ and $b$ living on a two-dimensional spacetime (with non-minimal couplings).

%%%%%%%%%%%%%%%%%%%%%%%%%%%%%%%%%%%%%%%%%%%%%%%%%%%%%
\section{Holographic Renormalization} \label{section:Holographic Renormalization}

In this section, we give the result of the holographic renormalization procedure in the usual Fefferman-Graham gauge \cite{Skenderis:2002wp}. This procedure consists in adding local, gauge invariant boundary counterterms to the on-shell action such that divergences cancel. The renormalized on-shell action is then interpreted as the generating functional of correlation functions of the boundary matrix theory. (The discussion in \cite{Skenderis:2002wp} is for Euclidean correlation functions. In Lorentzian signature, there are various correlation functions and extra care is needed; see for instance \cite{Son:2002sd, Herzog:2002pc, Skenderis:2008dh,Skenderis:2008dg}.)

We recall that we are now effectively working with a two-dimensional metric, whose Fefferman-Graham form is
\begin{align}
\label{44}
ds^2&=\frac{1}{z^2}\left(-f(t,z)dt^2+dz^2\right)\equiv\gamma(t,z)dt^2+\frac{dz^2}{z^2}.
\end{align}
Consider field fluctuations about a background,
\begin{align}
\label{74}
f(t,z)&=f_0(t,z)+\epsilon\ f_1(t,z),\\
\label{75}
\phi(t,z)&=\phi_0(t,z)+\epsilon\ \phi_1(t,z),\\
\label{76}
b(t,z)&=\frac{25}{4}+ \epsilon\ b_1(t,z),
\end{align}
where we can recognize the background value of the breathing mode from \eqref{51}. As mentioned in the Introduction, one cannot consider a dynamical breathing mode at the nonlinear level. Indeed, its presence strongly deforms the boundary away from $AdS_2$, as is manifest in the near-boundary asymptotic expansion of Appendix \ref{App:Asymptotics}. On the other hand, treating this mode perturbatively produces the holographic correlation functions and will be sufficient for our purposes. In the path integral formalism applied to matrix theory this amounts to considering sources in the generating functional which allows for the computation of $n$-point correlators by functional differentiation, and setting these to zero at the end of the calculation.

In order to perform holographic renormalization, we only need to know about the near-boundary asymptotics of these fields, which can be easily deduced by studying the equations of motions in the limit $z\rightarrow 0$. By expanding the fields in (possibly fractional) powers of $z$, one usually finds up to two undetermined coefficients per bulk field. The first one is typically the leading coefficient in that expansion and is usually referred to as the field \textit{source}. It needs to be specified as Dirichlet boundary value. The second undetermined coefficient is a subleading one and is directly related to the expectation value of the operator dual to the bulk field under consideration, in the corresponding matrix theory state. A necessary ingredient for computing correlation functions in a specific state is to know the solution deeper in the bulk (not only near the boundary), which fixes uniquely the subleading coefficient. We postpone this to Section~\ref{section:Propagators}, as it is not required for carrying out holographic renormalization.

In Appendix~\ref{App:Asymptotics} one can find details of the near-boundary asymptotics of the fields. The outcome is that there are sources for each of the physical fields considered here, but only the breathing mode has a subleading coefficient left undetermined by this asymptotic analysis. This comes from the fact that the metric and dilaton are non-propagating fields in this two-dimensional system, whereas the breathing mode does propagate.

As already said, holographic counterterms are boundary terms added to the action such that, when evaluated on-shell, the small $z$ cutoff (corresponding to a UV cutoff in the dual field theory) can be removed without generating boundary divergences. This means that the divergences coming from the on-shell action are to be cancelled against those coming from the counterterms. In Appendix~\ref{App:Onshell} we give the gauge invariant form of the on-shell action up to quadratic order in the expansion parameter $\epsilon$. In this work we are mainly interested in non-equal-time two-point correlation functions, and we will therefore only give the relevant boundary counterterms. These correlators encode non-trivial dynamical information about the dual matrix theory state. For the interested reader, we comment on one-point functions in Appendix \ref{App:One-point} and explain why the information they encode is somewhat trivial. Other counterterms are needed to ensure finiteness of the renormalized on-shell action, but these will not contribute to the values of one-point functions and non-equal-time two-point correlators.\footnote{For example, the full counterterm at order zero in $\epsilon$ is given by
\begin{equation*}
S^{count}=\frac{2 \pi^{9/2}}{\Gamma \left(\frac{9}{2}\right)}\left(\frac{25}{4}\right)^4\int dt\ \sqrt{-\gamma}\ e^{-\frac{6}{7}\phi} \left[-\frac{18}{5}+\frac{20}{49}\left( \partial \phi \right)^2-\frac{20}{21}\ \nabla^2 \phi \right],
\end{equation*}
where differential operators are boundary ones. Note that the first term is identical to \eqref{112} at order zero in $\epsilon$.} Neither should we care about terms local with respect to sources (those are often referred to as contact terms), since they contribute to equal-time correlators only. Here we therefore give the gauge invariant expression of the relevant boundary counterterms:
\begin{align}
\label{112}
S^{count}&=\frac{2 \pi^{9/2}}{\Gamma \left(\frac{9}{2}\right)}\int dt\ \sqrt{-\gamma}\ e^{-\frac{6}{7}\phi} \left[-\frac{5^7 .\ 23}{2^7}+\frac{3\ .\ 5^5 .\ 7}{2^3}\ b(t,z)-\frac{5^4 .\ 7}{2^2}\ b(t,z)^2\right].
\end{align}
One can check that leading divergences are thereby canceled against those present in the on-shell action \eqref{11} and that the finite renormalized on-shell action is given by
\begin{align}
\label{27}
S^{ren}&\equiv  S^{on-shell}+S^{count}\\
\label{116}
&=\frac{\pi^{9/2}}{\Gamma \left(\frac{9}{2}\right)}\int dt\ \sqrt{f_0^{(0)}}\ e^{-\frac{6}{7}\phi_0^{(0)}}\left[\frac{5^8}{2^7}\ M_0\ e^{\frac{4}{3}\phi_0^{(0)}}\right]\\
&+\epsilon\ \frac{\pi^{9/2}}{\Gamma \left(\frac{9}{2}\right)}\int dt\ \sqrt{f_0^{(0)}}\ e^{-\frac{6}{7}\phi_0^{(0)}}\left[\frac{5^{13}\ .\ 13}{2^{10} \ .\ 7^2}\ M_0^2\ e^{\frac{8}{3}\phi_0^{(0)}}\ b_1^{(-\frac{14}{5})}\right]\nonumber\\
&+\epsilon^2\ \frac{\pi^{9/2}}{\Gamma \left(\frac{9}{2}\right)}\int dt\ \sqrt{f_0^{(0)}}\ e^{-\frac{6}{7}\phi_0^{(0)}}\left[\frac{5^3 .\ 7^3}{2 \ .\ 3}\ b_1^{(-\frac{14}{5})}\ b_1^{(\frac{28}{5})}\right]+ \text{contact terms},\nonumber
\end{align}
where in the first line we implied a limit removing a small $z$ cutoff.
The final expression explicitly depends on the Fefferman-Graham expansion coefficients left undetermined by the asymptotic analysis, as well as on the mass parameter $M_0$ characterizing the most general solution to the gravity-dilaton system \eqref{70}-\eqref{71}; see Appendix~\ref{App:Asymptotics}. The leading terms in the asymptotical expansions, which are also called \textit{sources}, are $f_0^{(0)}, \phi_0^{(0)}$ and $b_1^{(-\frac{14}{5})}$. The only subleading undetermined coefficient $b_1^{(\frac{28}{5})}$ is related to the expectation value of the matrix theory operator $T^{--}$ as we will see in Section~\ref{section:Linear Response}. In the following sections, we will use the notation $b_1^{source}\equiv b_1^{(-\frac{14}{5})}$, hoping that this will make it easier for the reader to follow the logic of the computations. The renormalized action is to be understood as the generating functional of matrix theory, and will be used in Section~\ref{section:Linear Response} to compute non-equal-time correlators.

\newpage
%%%%%%%%%%%%%%%%%%%%%%%%%%%%%%%%%%%%%%%%%%%%%%%%%%%%%%%%% 
\section{Retarded Boundary-to-Bulk Propagators} \label{section:Propagators}
In this section we compute the \textit{retarded boundary-to-bulk propagators} of the linearized breathing mode fluctuations, for two different backgrounds. The first one is simply pure $AdS_2$. The second one is the thin shell limit \eqref{72}-\eqref{73} of the exact analytic solution \eqref{70}-\eqref{71} presented in the Introduction, whose dual representation can be thought of as matrix theory quenched from its ground state to a non-zero temperature state by sudden energy injection.

For convenience we recall here the background solution. The metric in ingoing Eddington-Finkelstein gauge adapted to the description of collapsing spacetimes is
\begin{equation}
\label{15}
ds^2=-\frac{1}{x^2}\left(2dvdx+g(v,x)dv^2\right).
\end{equation}
In this gauge, the most general background solution is given in closed form by
\begin{align}
\label{40}
\phi_0(v,x)&=\phi_0^{(0)}(v)+\frac{21}{10}\log x,\\
\label{16}
g_0(v,x)&=1+\frac{20}{21}\dot{\phi}_0^{(0)}(v)\ x-M_0\ e^{\frac{4}{3}\phi_0^{(0)}(v)}x^{14/5}.
\end{align}
We provide a derivation of this solution in Appendix~\ref{App:Background Solution}. The free function $\phi_0^{(0)}$ needs to be specified as Dirichlet condition and is interpreted as source for the scalar operator dual to the dilaton.

We then consider fluctuations about this general background solution, allowing for the breathing mode which, as mentioned earlier, is the only propagating degree of freedom in the reduced action \eqref{61}. Using the linearized equations of motion, one can derive a decoupled equation that this mode must satisfy,
\begin{multline}
\label{17}
-2x^2\ \dot{b_1}'(v,x)+\frac{9}{5}x\ \dot{b_1}(v,x)+\left(x^2+\frac{20}{21}\dot{\phi}_0^{(0)}(v)\ x^3-M_0\ e^{\frac{4}{3}\phi_0^{(0)}}x^{24/5}\right)b_1^{''}(v,x)\\
+\left(-\frac{9}{5}\ x+\frac{2}{21}\dot{\phi}_0^{(0)}(v)\ x^2-M_0\ e^{\frac{4}{3}\phi_0^{(0)}}x^{19/5}\right)b_1'(v,x)-\frac{392}{25}\ b_1(v,x)=0.
\end{multline}

A boundary-to-bulk propagator is a solution to this equation which reduces to a delta function profile at the boundary.
We now compute the retarded boundary-to-bulk propagator, which vanishes outside the future lightcone of the boundary point where the delta function has support, for both backgrounds under consideration.

%%%%%%%%%%%%%%%
\paragraph{Vacuum AdS}
The pure $AdS_2$ background solution is obtained by setting $M_0=0$ and  $\phi_0^{(0)}=0$ for all times in \eqref{40}-\eqref{16}. 
We derive the retarded boundary-to-bulk propagator (with delta function support at $t=0$ at the boundary)\footnote{Note that in pure $AdS_2$ the Eddington-Finkelstein coordinates can be simply related to the Fefferman-Graham ones: $x=z$ and $v=t-z$. For the general background solution \eqref{15}-\eqref{16} this is still valid asymptotically, $x\rightarrow z$ and $v \rightarrow t$ in the limit $z\rightarrow 0$.} in Appendix~\ref{App:Propagators}; its expression in Eddington-Finkelstein gauge is
\begin{align}
\label{68}
G_R^{AdS}(v,z)&=\frac{2\Gamma\left(\frac{47}{10}\right)}{\sqrt{\pi}\ \Gamma\left(\frac{21}{5}\right)}\ \sin\left(\frac{47\pi}{10}\right)\theta\left(v\right) \frac{z^{28/5}}{(v(v+2z))^{47/10}}.
\end{align}

Note that the ingoing lightlike coordinate $v$ reduces to the boundary time $t$ in the near boundary limit $z\rightarrow 0$. In that limit one can show that $G_R^{AdS}(t,z)\simeq z^{-14/5} \delta(t)$ which is the appropriate asymptotic behavior for a delta source at the boundary; see Appendix~\ref{App:Asymptotics}. Therefore a generic source profile $b_1^{source}(t)$ induces a bulk solution $b_1(t,z)$ in its future lightcone:
\begin{align}
\label{14}
&b_1(t,z)=\int_{-\infty}^{\infty}dt'\ G_R^{AdS}(t-t',z)\ b_1^{source}(t'),\\
&\lim\limits_{z\rightarrow 0}b_1(t,z)\simeq z^{-14/5}b_1^{source}(t).
\end{align}

%%%%%%%%%%%%%%%
\paragraph{Thin Shell Collapse}
Assuming a fixed non-zero mass parameter $M_0$, the choice of a time-dependent dilaton source profile $\phi_0^{(0)}$ as Dirichlet boundary condition offers the possibility of having collapsing solutions. Here we consider a thin shell limit in which vacuum AdS is glued to a black hole background at a null surface $v=v_0$. This is achieved by choosing the following profile for the dilaton source:
\begin{align}
\label{67}
&v<v_0: \qquad \begin{cases}
\phi_0^{(0)}(v)\rightarrow-\infty,\\
\dot{\phi}_0^{(0)}(v)=0,
\end{cases}\\
&\nonumber\\
&v>v_0: \qquad \phi_0^{(0)}(v)=0.\label{67a}
\end{align}
Note that we can also use any dilaton source profile after the gluing surface $v=v_0$, resulting in a generic time-dependent black hole background. Note also that the limit \eqref{67} is perfectly acceptable since one expects supergravity to be valid for small value of $e^{\phi}$ (except if one were to consider observables with strong support in the deep infrared regime, as discussed in Section~\ref{section:Review}).

Consider a delta function source $b_1^{source}(t)=\delta(t-t_s)$ on the boundary before the shell collapse, at $t_s<v_0$. The retarded boundary-to-bulk propagator on the pure AdS segment is already known,
\begin{equation}
G_R^{Shell}(v,t_s;z)=\frac{2\Gamma\left(\frac{47}{10}\right)}{\sqrt{\pi}\ \Gamma\left(\frac{21}{5}\right)}\ \sin\left(\frac{47\pi}{10}\right)\theta\left(v-t_s\right) \frac{z^{28/5}}{((v-t_s)(v-t_s+2z))^{47/10}}, \quad v<v_0,
\end{equation}
and one simply has to numerically continue it across the shell located at $v=v_0$. Therefore, by using $G_R^{Shell}(v_0,t_s;z)$ at this surface as initial condition for solving the equation of motion \eqref{17}, one can extend the retarded propagator to the black hole part of the geometry. For this we used the \verb|NDSolve| command of \emph{Mathematica}, with asymptotic boundary condition $b_1(v,x=0)=0$. Once numerical evolution of the retarded boundary-to-bulk propagator is obtained, one can extract the asymptotic coefficient that was left  unfixed by asymptotic analysis for Dirichlet boundary condition. This coefficient is the one appearing at order $z^{28/5}$ near the boundary; see Appendix~\ref{App:Asymptotics} for more details. In particular, in the limit $z \rightarrow 0$ and for non-equal times one has
\begin{equation}
G_{R}(v,t_s;z)=G_{R}^{(\frac{28}{5})}(t,t_s)z^{28/5}+\mathcal{O}(z^{33/5}), \qquad t\neq t_s,
\end{equation}
and the leading coefficient $G_{R}^{(\frac{28}{5})}$ is directly related to the retarded two-point function of the matrix theory dual operator, as we will soon see. This coefficient as well as subleading ones can be extracted from the numerical solution by fitting a power series of the form \eqref{85} appropriately expressed in coordinates $(v,x)$, close to the boundary $x \rightarrow 0$. Note that the first three towers of terms in the expansion \eqref{85} are effectively missing whenever the breathing mode source is turned off, which is presently the case for $t\neq t_s$.

\begin{figure}
	\begin{center}
		\includegraphics[scale=0.6]{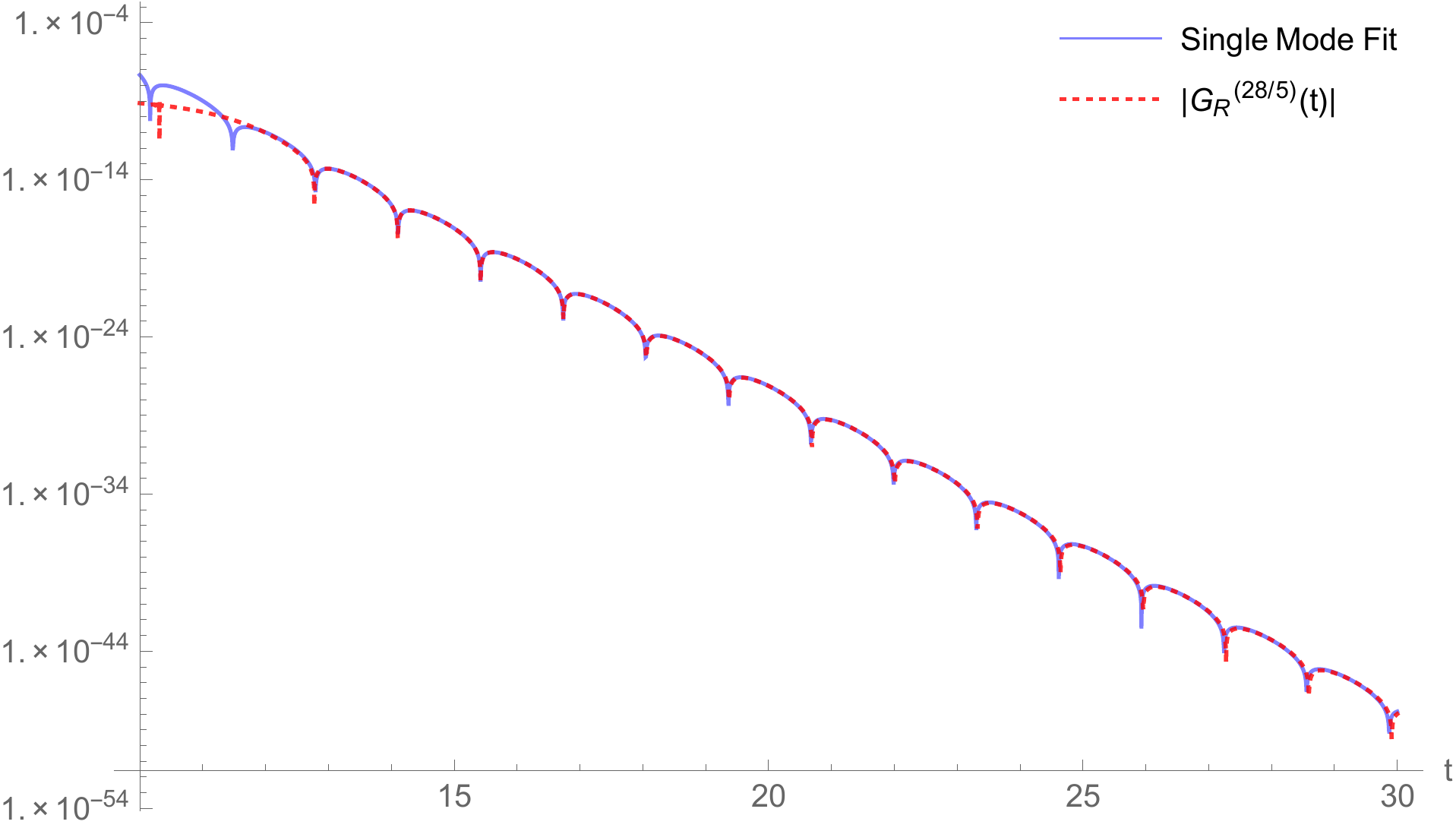}
		\caption{Logarithmic plot of $ \vert G_{R}^{(\frac{28}{5})}(t) \vert $ (dotted red line) and fitted single oscillating and decaying mode (blue line). The black hole mass is chosen to be $M_0=0.5$ and the delta source is located at $t_s=0$. The $t$-axis starts at the gluing surface $v_0=10$.}
		\label{Single Mode Fit}
	\end{center}
\end{figure}

The main result coming out of the numerical evolution from the gluing surface at $v=v_0$ is that this leading coefficient is very rapidly dominated by a single oscillating and decaying mode of the form
\begin{equation}
\label{79}
G_{R}^{(\frac{28}{5})}(t,t_s)\sim e^{-i \omega_R t}e^{-\omega_I t}, \qquad t>v_0.
\end{equation}
The more massive the black hole, the more rapidly this mode dominates. The timescale for perfect matching (within numerical errors) is of the order of the inverse horizon radius $r_h=M_0^{5/14}$ of the associated final black hole.  We show an example of this behavior in Figure \ref{Single Mode Fit}. 
In Figure \ref{Extracted frequencies}, we give the extracted values of $\omega_R$ and $\omega_I$ for various black hole masses.
\begin{figure}
	\begin{center}
		\begin{tabular}{ |c|c|c| } 
			\hline
			$M_0$ & $\omega_R$ & $\omega_I$ \\ 
			\hline
			1 & 5.9566 & 3.05939 \\
			0.5 & 4.64025 & 2.39403 \\
			0.01 & 1.14375 & 0.586577 \\
			0.03 & 1.70325 & 0.871438 \\
			0.05 & 2.04138 & 1.04437 \\
			0.07 & 2.29722 & 1.18774 \\
			0.1 & 2.61204 & 1.34623 \\
			\hline
		\end{tabular}
		\caption{Extracted frequencies from $G_{R}^{(\frac{28}{5})}(t,t_s)$ for various values of the black hole mass $M_0$.}
		\label{Extracted frequencies}
	\end{center}
\end{figure} 
This leads to the following frequency-to-temperature ratios:

\begin{align}
\label{77}
&\frac{\omega_R}{T_H}=13.7(3),\\
\label{78}
&\frac{\omega_I}{T_H}=26.7(0),
\end{align}
where the Hawking-Unruh black hole temperature is given by \cite{Matsuo:2013jda}
\begin{align}
\label{81}
T_H=\frac{7}{10 \pi}r_h.
\end{align}

Importantly, the single complex frequency $\omega=\omega_R-i \omega_I$ for a choice of final black hole temperature $T_H$ can be matched to the lowest quasinormal frequency of breathing mode fluctuations around a static black hole at the same temperature. Indeed, using a numerical shooting method we compute in Appendix~\ref{App:LowestQNM} the first and second quasinormal frequencies whose values are given by
\begin{align}
\label{36}
\text{First QNM:} \qquad &\frac{\omega_R}{T_H}=13.7,\\
\label{37}
&\frac{\omega_I}{T_H}=26.7,
\end{align}
\begin{align}
\label{107}
\text{Second QNM:} \qquad &\frac{\omega_R}{T_H}=18.6,\\
&\frac{\omega_I}{T_H}=36.6.\label{107a}
\end{align}

Similar results in a related context have been found in \cite{Balasubramanian:2012tu,David:2015xqa}. In Appendix~\ref{App:LowestQNM} we also derive the universality of the frequency-to-temperature ratio, implicitly indicated in \eqref{77}-\eqref{107a}.

%%%%%%%%%%%%%%%%%%%%%%%%%%%%%%%%%%%%%%%%%%%%%%%%%%%%%%%%%
\section{Linear Response in Matrix Theory} \label{section:Linear Response}
We are now ready to compute non-equal-time retarded two-point functions of the matrix theory operator $T^{--}$, as probe of thermalization. We assume a $\delta$-function source profile $b_1^{source}(t)=\delta(t-t_s)$ on the boundary and recall the relation \eqref{14} between this source and the induced retarded bulk solution. We note that the field dual to $T^{--}$ is naturally given by $e^{-\frac{6}{7} \phi} b$ rather than simply $b$. Indeed, couplings of the boundary operators to their dual bulk fields can be recovered from considering the effective D-brane action in a given bulk field background. For example, at the level of a single D-particle (matrix theory being the low-energy limit of the $U(N)$ generalization of this action), the metric-dependent part of the action in the dual frame defined by \eqref{55} is given by
	\begin{equation}
	S_{dual}^{DBI}\sim -\int dt\ e^{-\frac{6}{7} \phi} \sqrt{-\dot{x}^M \dot{x}^N g_{MN}}.
	\end{equation}
	Linearizing around the Minkowski background,
	\begin{align}
	g_{MN}&=\eta_{MN}+h_{MN},\\
	\phi&=\phi_0+\phi_1,
	\end{align}
	one gets
	\begin{equation}
	S_{dual}^{DBI}\sim - \int dt\ e^{-\frac{6}{7}\phi_0}\left[ (1-v^2)^{1/2}(1-\frac{6}{7}\phi_1)-\frac{1}{2}(1-v^2)^{-1/2} \left(h_{00}+2h_{0i}\ v^i+h_{ij}\ v^iv^j\right)\right],
	\end{equation}
	where we have defined $v^i\equiv \dot{x}^i$. One can therefore conclude that fields coupling to operators like $v^{i}v^{j}$ or $v^4$ (the latter being proportional to $T^{--}$ in this $U(1)$ case) are $e^{-\frac{6}{7} \phi_0} h$ rather than the simple metric perturbations $h$. (Of course, if the background dilaton $\phi_0$ is time-independent, this distinction becomes unimportant.)
	
According to the usual holographic dictionary \cite{Skenderis:2002wp} and focusing on the order $\epsilon^2$ piece of the renormalized on-shell action \eqref{116} that will lead to a non-equal-time correlator, we then have
\begin{align}
\langle \mathcal{O}(t)\rangle_s&=\frac{1}{\sqrt{f_0^{(0)}}}\frac{\delta S^{ren}}{\delta\ \epsilon e^{-\frac{6}{7} \phi_0^{(0)}}b_1^{source}}\\
&=\epsilon\ \frac{\pi^{9/2}}{\Gamma \left(\frac{9}{2}\right)}\ \frac{5^3 .\ 7^3}{2\ .\ 3}\ b_1^{(\frac{28}{5})}(t)+...\ ,\\
&=\epsilon\ \frac{\pi^{9/2}}{\Gamma \left(\frac{9}{2}\right)}\ \frac{5^3 .\ 7^3}{2\ .\ 3}\ \int dt'\ G_{R}^{(\frac{28}{5})}(t,t')\ b_1^{source}(t')+\ldots\ ,
\end{align}
where $\ldots$ stands for terms analytic in $b_1^{source}(t)$ which only contribute to equal-time two-point functions. Since the one-point function in absence of source is trivial, $\langle \mathcal{O}(t)\rangle_{s=0}=0$, the response function is simply 
\begin{equation}
\delta \langle \mathcal{O}(t)\rangle= \langle \mathcal{O}(t)\rangle_s\equiv\int dt' \ \Delta_R(t,t')\ e^{-\frac{6}{7} \phi_0^{(0)}(t')}\ b_1^{source}(t'),
\end{equation}
and we identify the non-equal-time retarded two-point function of the matrix theory as
\begin{align}
\label{28}
\Delta_R(t,t')
&= \epsilon\ \frac{\pi^{9/2}}{\Gamma \left(\frac{9}{2}\right)}\ \frac{5^3 .\ 7^3}{2\ .\ 3}\ e^{\frac{6}{7}\phi_0^{(0)}(t')}\ G_{R}^{(\frac{28}{5})}(t,t'), \qquad t\neq t'.
\end{align}

%%%%%%%%%%%%%%
\paragraph{Ground state Two-Point Function}
As a first special case we consider a pure $AdS_2$ background, which is dual to the matrix theory ground state. The retarded boundary-to-bulk propagator is given in \eqref{68} while the dilaton source is $\phi_0^{(0)}=0$, such that one finds 
\begin{align}
\Delta_R^{ground}(t,t_s)
&=\epsilon\ \frac{\pi^{9/2}}{\Gamma \left(\frac{9}{2}\right)}\ \frac{5^3 .\ 7^3}{2\ .\ 3}\ G_{R}^{(\frac{28}{5})}(t,t_s),\\
&=\epsilon\ \frac{5^3 .\ 7^3}{3}\ \pi^{4}\frac{\Gamma\left(\frac{47}{10}\right)}{\Gamma\left(\frac{9}{2}\right)\Gamma\left(\frac{21}{5}\right)}\ \sin\left(\frac{47\pi}{10}\right)\theta\left(t-t_s\right) \frac{1}{(t-t_s)^{47/5}}, \qquad t\neq t_s.
\end{align}
One can check that the power decay agrees with previous work \cite{Sekino:1999av}.

%%%%%%%%%%%%%%%
\paragraph{Thermalizing Two-Point Function} 
The collapsing thin shell setup that we studied in Section \ref{section:Propagators} is dual to matrix theory initially in its ground state and suddenly excited by energy injection. The retarded thermalizing two-point function in this quenched state with insertion times $t_s < t_{quench}\equiv v_0 < t$ is rapidly dominated by a single complex frequency for $t>t_{quench}$, as analyzed in Section \ref{section:Propagators}. Its expression is given by 
\begin{equation}
\label{80}
\Delta_R^{thermalizing}(t,t_s)=\epsilon\ \frac{\pi^{9/2}}{\Gamma \left(\frac{9}{2}\right)}\ \frac{5^3 .\ 7^3}{2 \ .\ 3}\ G_{R}^{(\frac{28}{5})}(t,t_s), \quad t_s<t_{quench}< t,
\end{equation}
where $G_{R}^{(\frac{28}{5})}(t,t_s)$ can be studied using the single complex frequency given in \eqref{77}-\eqref{78} and \eqref{36}-\eqref{37}.

\begin{figure}
	\centering
	\begin{subfigure}[b]{.5\textwidth}
		\centering
		\includegraphics[width=1\linewidth]{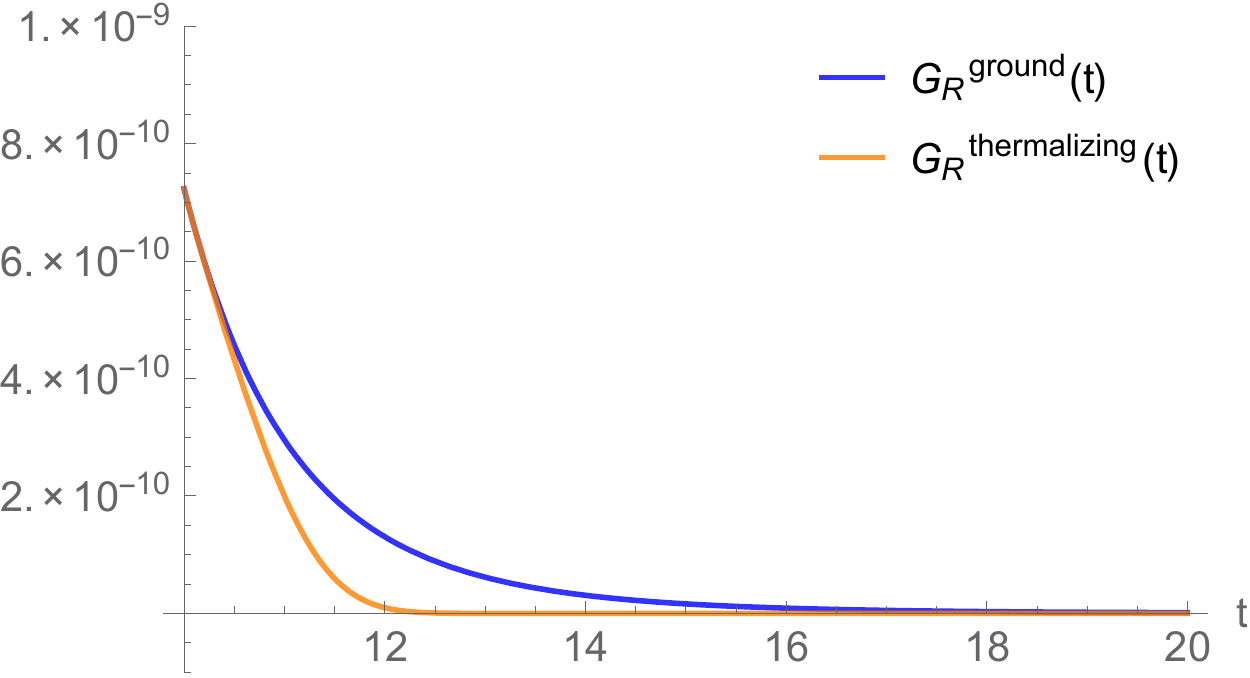}
		\caption{}
		\label{fig:sub1}
	\end{subfigure}%
	\begin{subfigure}[b]{.5\textwidth}
		\centering
		\includegraphics[width=1\linewidth]{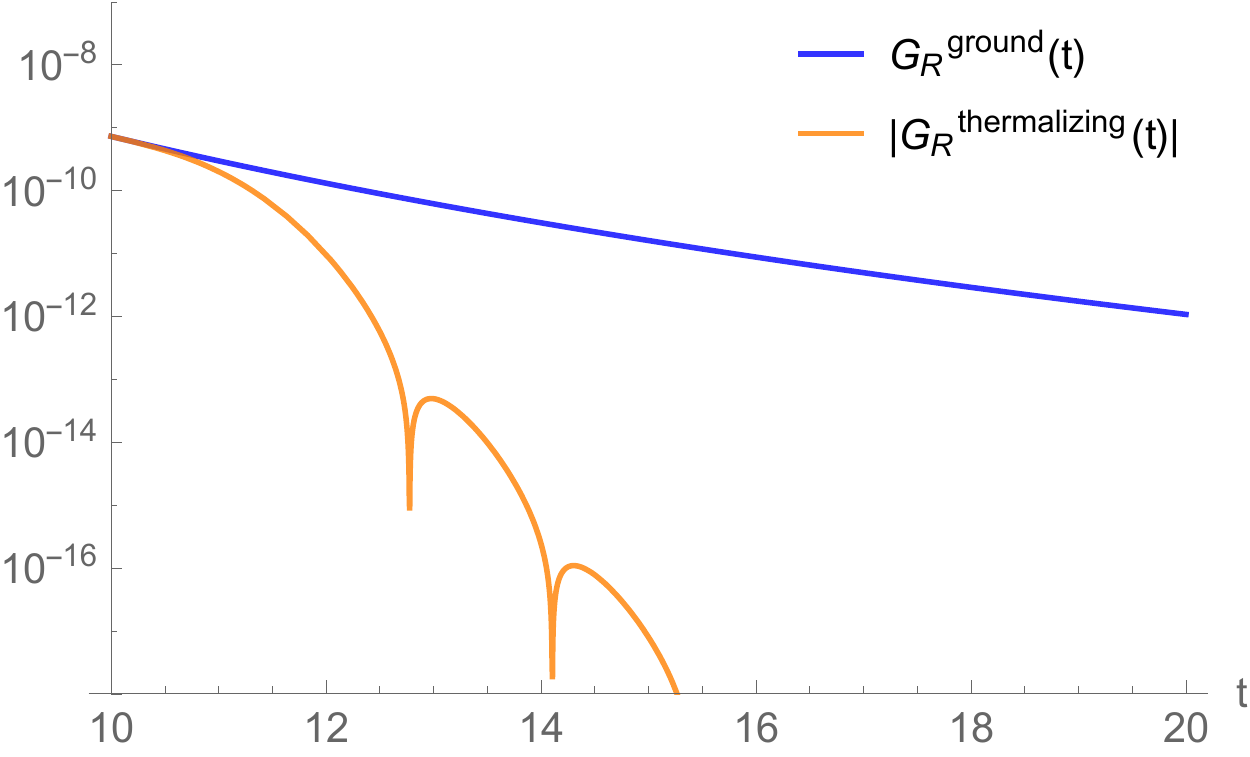}
		\caption{}
		\label{fig:sub2}
	\end{subfigure}
	\caption{Comparison of the ground state and thermalizing retarded two-point functions of the matrix theory operator $T^{--}$, in terms of the function $G_R^{(\frac{28}{5})}$. For this computation, the black hole mass is fixed to $M_0=0.5$ and the delta source is located at $t_s=0$. The $t$-axis starts at the shell $t_{quench}=10$. (a) Linear scale. (b) Logarithmic scale.}
	\label{Comparison}
\end{figure}

In Figure \ref{Comparison}, we compare thermalizing and ground state two-point functions. One can see that the former relaxes more rapidly, as can be generally expected from the exponential decay of quasinormal modes. (See also equation \eqref{79} for the generic behavior.) Of course, this is consistent with the general picture of how thermalizing observables should evolve.

%%%%%%%%%%%%%%%%%%%%%%%%%%%%%%%%%%%%%%%%%%%%%%%%%%%%%%%%
\section*{Acknowledgments}

We would like to thank A.~Castro, M.~Hanada, D.~Hofman, I.~Papadimitriou and K.~Van Acoleyen for useful discussions.
The work of B.C.\ and K.N.\ is supported in part by the Belgian Federal Science Policy Office through the Interuniversity Attraction Pole P7/37, by FWO-Vlaanderen through projects G020714N and G044016N, and by Vrije Universiteit Brussel through the Strategic Research Program ``High-Energy Physics''. The work of O.E.\ is funded under CUniverse research promotion project by Chulalongkorn University (grant reference CUAASC). K.N.\ would like to thank UCLouvain for funding, as well as the CP3 Centre for hospitality during the course of this work. B.C.\ thanks the organizers of the Nordita program ``Black Holes and Emergent Spacetime'' for hospitality while this work was in progress.

%%%%%%%%%%%%%%%%%%%%%%%%%%%%%%%%%%%%%%%%%%%%%%%%%%%%%%%%%%
\appendix
\section{Details of the Dimensional Reduction} \label{App:Reduction}
In this appendix we work out the details leading to the effective two-dimensional action \eqref{61}. We also show how one gets the associated equations of motions. With the assumptions \eqref{62},\eqref{48},\eqref{63} of Section \ref{section:Dimensional Reduction}, equations \eqref{1}-\eqref{2} impose 
\begin{equation}
\label{42}
A(x^M)=A(x^\mu)=\sqrt{2C}\ e^{-\frac{6}{7} \phi} b^{-4},
\end{equation}
with $C$ a constant. A simple way to identify the value of $C$ that supports asymptotically $AdS_2 \times S^8$ solutions of \eqref{3}-\eqref{2} is by computing 
\begin{equation}
\label{10}
F^2=-2|A(x^M)|^2=-4C\ e^{-\frac{12}{7}\phi} b^{-8}.
\end{equation}
Any asymptotically AdS solution must agree in the asymptotic region with \eqref{51}- \eqref{53}, which implies that
\begin{equation}
C=\frac{1}{2}\left(\frac{14}{5}\right)^2 \left(\frac{25}{4}\right)^8.
\end{equation}
The dilaton equation of motion \eqref{4} then becomes
\begin{equation}
\label{5}
R-\frac{16}{49} (\partial \phi)^2+\frac{16}{21} \nabla^2\phi -C b^{-8}=0.
\end{equation}
For the metric equation of motion \eqref{3}, one can consider three relevant cases: those with $(M,N)=(\mu,\nu)$, $(M,N)=(i,j)$ and $(M,N)=(\mu,i)$. One can show that the first one gives the set of equations 
\begin{equation}
\label{6}
-R_{\mu\nu}+\frac{20}{49}\partial_\mu\phi \partial_\nu\phi-\frac{6}{7}\nabla_\mu\partial_\nu\phi
+\frac{1}{2}g_{\mu\nu}\left(R-\frac{8}{7}(\partial\phi)^2+\frac{12}{7} \nabla^2\phi-Cb^{-8}\right)=0,
\end{equation}
that the second one gives the unique equation
\begin{equation}
\label{7}
-\frac{1}{8}g^{kl}R_{kl}-\frac{3}{7b}g^{\mu\nu}\partial_{\mu}b \partial_{\nu}\phi+\frac{1}{2}\left(R-\frac{8}{7}(\partial\phi)^2+\frac{12}{7} \nabla^2\phi+Cb^{-8}\right)=0,
\end{equation}
while the third one is trivially satisfied.
 
Instead of working with the fields $\{g_{MN},\phi\}$, one can work with $\{g_{\mu\nu}, b, \phi\}$. First one notices that all quantities in the previous equations can be written in terms of $\{g_{\mu\nu}, b, \phi\}$. In particular, we have
\begin{align}
\label{30}
K&=\tensor[^{(2)}]{K}{}+4 b^{-1} n^\mu \partial_\mu b,\\
\label{82}
R&=\tensor[^{(2)}]{R}{}+56 b^{-1}-10b^{-2}(\partial b)^2-8b^{-1}\ \tensor[^{(2)}]{\nabla}{^2} b,\\
\label{65}
R_{\mu\nu}&=\tensor[^{(2)}]{R}{_\mu_\nu}+2b^{-2}\partial_\mu b\ \partial_\nu b-4b^{-1}\ \tensor[^{(2)}]{\nabla}{_\mu}\partial_\nu b,\\
\nabla^2 \phi&=\tensor[^{(2)}]{\nabla}{^2}\phi+4b^{-1}\partial_\mu b\ \partial^\mu \phi,\\
\label{66}
\nabla_\mu\partial_\nu\phi&=\tensor[^{(2)}]{\nabla}{_\mu}\partial_\nu\phi,\\
\label{93}
\frac{1}{8}g^{kl}R_{kl}&=\left(7b^{-1}-\frac{1}{2}\ b^{-1}\ \tensor[^{(2)}]{\nabla}{^2} b-\frac{3}{2}\ b^{-2}(\partial b)^2\right),
\end{align}
where $ ^{(2)}  $ indicates that all geometric quantities are constructed from the two-dimensional metric $g_{\mu\nu}$ only, and where $n^\mu$ is the outward-pointing normal vector to the timelike boundary. These relations can be inserted in \eqref{5}-\eqref{7} in order to get the effective two-dimensional equations of motions.

Using \eqref{10},\eqref{30},\eqref{82} together with $\sqrt{-g}=\sqrt{-\tensor[^{(2)}]{g}{}}\sqrt{g_{S^8}}\ b^4$, one can show that the action \eqref{60} becomes
\begin{align}
\label{92}
S_{eff}=&\frac{2\pi^{9/2}}{\Gamma \left(\frac{9}{2}\right)}\int\!\! d^{2}x \sqrt{\!-\tensor[^{(2)}]{g}{}}\, e^{-\frac{6}{7}\phi} \left[b^4\, \tensor[^{(2)}]{R}{}+56 b^{3}+14b^{2}(\partial b)^2-\frac{48}{7} b^{3}\partial_{\mu}b\, \partial^{\mu}\phi+\frac{16}{49}b^{4} (\partial \phi)^2-\frac{C}{b^4}\right]\nonumber\\
&+\frac{4\pi^{9/2}}{\Gamma \left(\frac{9}{2}\right)}\int d^{1}x\ \sqrt{-\tensor[^{(2)}]{\gamma}{}}\ e^{-\frac{6}{7}\phi}b^4 \ \tensor[^{(2)}]{K}{}.
\end{align}

Note that the bulk term $-Cb^{-4}$ in the effective action is obtained not only from the term $-\frac{1}{4}e^{\frac{6}{7}\phi}F^2$ in the bulk part of the action \eqref{60}, but also from the term $\frac{1}{2}e^{\frac{6}{7}\phi}F^2$ in the boundary part of the action $S_{extra}$ \eqref{108}. One can further check that this gives the correct effective EOM's by extremization of $S_{eff}$. Although $S_{extra}$ is explicitly written as a Maxwell bulk term in \eqref{109} for fields configurations satisfying the gauge field EOM \eqref{1}, one would usually think that boundary terms should not contribute to the EOM's. In this case, there is an underlying reason why it does contribute. Indeed, fixing the gauge
\begin{align}
ds^2&=\frac{1}{z^2}\left(-f(t,z)dt^2+dz^2\right)+b(t,z)d\Omega_8^2,\\
C_z&=0,
\end{align} 
we deduce from \eqref{48}-\eqref{42} that
\begin{equation}
\label{110}
F_{0z}=\sqrt{2C}\sqrt{-\tensor[^{(2)}]{g}{}}\ e^{-\frac{6}{7}\phi}b^{-4}.
\end{equation}
A possible gauge potential is therefore given by
\begin{equation}
\label{111}
C_0(t,z)=-\sqrt{2C}\int_{\infty}^{z}dz'\sqrt{-\tensor[^{(2)}]{g(t,z')}{}}\ e^{-\frac{6}{7}\phi(t,z')}b(t,z')^{-4},
\end{equation}
which is a non-local functional of the fields $\tensor[^{(2)}]{g}{_\mu_\nu}, b, \phi$. In particular bulk variations of those fields induce boundary variations of $C_0$. Therefore, in order to get the EOM's for those fields by action extremization, it is not sufficient to consider variations of $\tensor[^{(2)}]{g}{_\mu_\nu}, b, \phi$ and $C_0$ localized in the bulk. An alternative way of seeing this is that for the subclass of fields configurations \eqref{110}-\eqref{111}, one can show that the boundary term $S_{extra}$ effectively produces a bulk term through the non-locality of $C_0$:
\begin{align}
S_{extra}&=\int d^9x \sqrt{-\gamma}\ n^{M} \left[e^{\frac{6}{7}\phi} F_{MN}C^{N}\right]\bigg\vert_{z=0}=-\frac{2\pi^{9/2}}{\Gamma \left(\frac{9}{2}\right)} \sqrt{2C} \int dt\ C_0(t,z=0)\\
&=\frac{4C\pi^{9/2}}{\Gamma \left(\frac{9}{2}\right)} \int dt \int_{\infty}^{0} dz \sqrt{-\tensor[^{(2)}]{g}{}}\ e^{-\frac{6}{7}\phi}b^{-4}=-\frac{4C\pi^{9/2}}{\Gamma \left(\frac{9}{2}\right)} \int d^2x \sqrt{-\tensor[^{(2)}]{g}{}}\ e^{-\frac{6}{7}\phi}b^{-4},
\end{align}
where $n^{M}=-z\delta^{M}_z$ is the outward-pointing unit normal vector to the boundary, and $\gamma$ its induced metric. The same result is of course directly obtained from the expression of $S_{extra}$ given in bulk form in \eqref{109}.

In the main body of this paper, we will exclusively use the two-dimensional effective system. We therefore omit the $ ^{(2)} $ superscript there, since all geometrical quantities will refer to the two-dimensional metric.

%%%%%%%%%%%%%%%%%%%%%%%%%%%%%%%%%%%%%%%%%%%%%%%%%%%%%%%%%%
\section{Derivation of General Background Solution} \label{App:Background Solution}
We provide a derivation of the general background solution \eqref{40}-\eqref{16} 
in ingoing Eddington-Finkelstein gauge
\begin{equation}
ds^2=-\frac{1}{x^2}\left(2dvdx+g(v,x)dv^2\right).
\end{equation}
\newline
In these coordinates, the equations of motion for the background fields are given by
\begin{align}
\label{102}
&\frac{16}{21} x^2 g^{(0,1)}(v,x) \phi ^{(0,1)}(v,x)-x^2 g^{(0,2)}(v,x)+2 x g^{(0,1)}(v,x)-\frac{16}{49} x^2 g(v,x) \phi^{(0,1)}(v,x)^2\\
&+\frac{16}{21} x^2 g(v,x) \phi^{(0,2)}(v,x)-2 g(v,x)+\frac{32}{49} x^2 \phi ^{(0,1)}(v,x) \phi^{(1,0)}(v,x)-\frac{32}{21} x^2 \phi ^{(1,1)}(v,x)+\frac{126}{25}=0,\nonumber\\
&\nonumber\\
\label{103}
&-\frac{3}{7} g^{(0,1)}(v,x) g(v,x) \phi ^{(0,1)}(v,x)+\frac{3}{7} g^{(1,0)}(v,x) \phi^{(0,1)}(v,x)-\frac{3}{7} g^{(0,1)}(v,x) \phi^{(1,0)}(v,x)\\
&-\frac{63 g(v,x)}{25 x^2}+\frac{4}{7} g(v,x)^2 \phi ^{(0,1)}(v,x)^2-\frac{6 g(v,x)^2 \phi ^{(0,1)}(v,x)}{7x}-\frac{6}{7} g(v,x)^2 \phi ^{(0,2)}(v,x)\nonumber\\
&-\frac{8}{7} g(v,x) \phi ^{(0,1)}(v,x) \phi ^{(1,0)}(v,x)+\frac{6 g(v,x) \phi^{(1,0)}(v,x)}{7 x}+\frac{12}{7} g(v,x) \phi ^{(1,1)}(v,x)\nonumber\\
&+\frac{20}{49} \phi^{(1,0)}(v,x)^2-\frac{6}{7} \phi ^{(2,0)}(v,x)=0,\nonumber\\
&\nonumber\\
\label{104}
&-\frac{3}{7} g^{(0,1)}(v,x) \phi ^{(0,1)}(v,x)+\frac{4}{7} g(v,x) \phi ^{(0,1)}(v,x)^2-\frac{6 g(v,x) \phi ^{(0,1)}(v,x)}{7x}\\
&-\frac{6}{7} g(v,x) \phi ^{(0,2)}(v,x)-\frac{36}{49} \phi^{(1,0)}(v,x) \phi ^{(0,1)}(v,x)+\frac{6}{7} \phi^{(1,1)}(v,x)-\frac{63}{25 x^2}=0,\nonumber\\
&\nonumber\\
\label{105}
&\frac{20}{49} \phi ^{(0,1)}(v,x)^2-\frac{12 \phi ^{(0,1)}(v,x)}{7 x}-\frac{6}{7} \phi ^{(0,2)}(v,x)=0,\\
&\nonumber\\
\label{106}
&\frac{6}{7} x^2 g^{(0,1)}(v,x) \phi ^{(0,1)}(v,x)-\frac{1}{2} x^2 g^{(0,2)}(v,x)+x g^{(0,1)}(v,x)-\frac{4}{7} x^2 g(v,x) \phi^{(0,1)}(v,x)^2\\
&+\frac{6}{7} x^2 g(v,x) \phi^{(0,2)}(v,x)-g(v,x)+\frac{8}{7} x^2 \phi ^{(0,1)}(v,x) \phi ^{(1,0)}(v,x)-\frac{12}{7}x^2 \phi ^{(1,1)}(v,x)+\frac{133}{25}=0.\nonumber
\end{align}
Integrating \eqref{105} readily gives 
\begin{equation}
\phi(v,x)=\phi^{(0)}(v)-\frac{21}{10}\log \left(\frac{1}{x}-A(v)\right),
\end{equation}
where $\phi^{(0)}(v)$ and $A(v)$ are arbitrary functions of $v$. Knowing $\phi(v,x)$, equation \eqref{102} is just an ordinary differential equation for $g$ with variable $x$. The solution is
\begin{align}
g(v,x)&=1+\left(\frac{20}{21}\dot{\phi}^{(0)}(v) -2 A(v)\right)x+\left(A(v)^2+2\dot{A}(v)-\frac{20}{21}A(v)\dot{\phi}^{(0)}(v)\right)x^2\\
&\qquad +\frac{C_1(v)\ x^{9/5}}{\left(1-A(v) x\right)^{4/5}}-\frac{C_2(v)\ x^{14/5}}{\left(1-A(v) x\right)^{4/5}}.
\end{align}      
The two functions $C_1(v)$ and $C_2(v)$ are then constrained by \eqref{103}-\eqref{104},
\begin{equation}
C_1(v)=0, \qquad C_2(v)=M_0\ e^{\frac{4}{3}\phi^{(0)}(v)},
\end{equation}
with $M_0$ a free \textit{mass parameter}. One can check that \eqref{106} is also satisfied. The function $A(v)$  can be gauged away through the replacement 
\begin{equation}
\frac{1}{\tilde{x}}=\frac{1}{x}-A(v),
\end{equation}
bringing the solution to the form \eqref{40}-\eqref{16},
\begin{align}
ds^2_{dual}&=-\frac{1}{\tilde{x}^2}\left[2\ dvd\tilde{x}+\left(1+\frac{20}{21}\dot{\phi}^{(0)}(v)\ \tilde{x}-M_0\ e^{\frac{4}{3}\phi^{(0)}(v)}\tilde{x}^{14/5}\right)dv^2\right],\\
\phi(v,\tilde{x})&=\phi^{(0)}(v)+\frac{21}{10}\log \tilde{x}.\nonumber
\end{align}

%%%%%%%%%%%%%%%%%%%%%%%%%%%%%%%%%%%%%%%%%%%%%%%%%%%%
\section{Bulk Field Asymptotics} \label{App:Asymptotics}
In this appendix we derive the near-boundary asymptotics of the bulk fields \eqref{74}-\eqref{76}, by analyzing the equations of motions in Fefferman-Graham gauge \eqref{44} in the limit $z \rightarrow 0$.

We start with the background fields. For illustrative purposes, we display their equations of motion, which can be derived by variation of \eqref{92} or by use of \eqref{5}-\eqref{66} together with $b(t,z)=\frac{25}{4}$:\footnote{If one needed to know the equations of motion involving also a dynamical breathing mode in Fefferman-Graham coordinates, it would be simpler to start with the covariant expressions \eqref{5}-\eqref{93} and use some code to generate the expressions in that gauge rather than copy huge formulas that we refrain from displaying here. See for example the useful \textit{TensoriaCalc} package for \textit{Mathematica} \cite{TensoriaCalc}.}

%Dilaton:
\begin{align}
\label{94}
&\frac{8}{3} z^2 f(t,z) f^{(0,1)}(t,z) \phi ^{(0,1)}(t,z)+\frac{8}{3} z^2 f^{(1,0)}(t,z) \phi ^{(1,0)}(t,z)+\frac{7}{2} z^2 f^{(0,1)}(t,z)^2\\
&-7 z^2 f(t,z) f^{(0,2)}(t,z)+7 z f(t,z) f^{(0,1)}(t,z)-\frac{16}{7} z^2 f(t,z)^2 \phi^{(0,1)}(t,z)^2 \nonumber\\
&+\frac{16}{7} z^2 f(t,z) \phi^{(1,0)}(t,z)^2+\frac{16}{3} z^2 f(t,z)^2 \phi^{(0,2)}(t,z)-\frac{16}{3} z^2 f(t,z) \phi ^{(2,0)}(t,z)+\frac{532}{25} f(t,z)^2=0,\nonumber\\
&\nonumber\\
%TT Component:
\label{95}
&4 z^2 f(t,z) \phi ^{(0,1)}(t,z)^2-6 z^2 f(t,z) \phi ^{(0,2)}(t,z)-6 z f(t,z) \phi ^{(0,1)}(t,z)-\frac{441}{25} f(t,z)\nonumber\\
&-\frac{8}{7} z^2 \phi^{(1,0)}(t,z)^2=0,\\
&\nonumber\\
%TZ Component:
\label{96}
&3 f^{(0,1)}(t,z) \phi ^{(1,0)}(t,z)-\frac{6 f(t,z) \phi ^{(1,0)}(t,z)}{z}+\frac{20}{7} f(t,z) \phi ^{(0,1)}(t,z) \phi ^{(1,0)}(t,z)\nonumber\\
&-6 f(t,z) \phi^{(1,1)}(t,z)=0,\\
&\nonumber\\
%ZZ Component:
\label{97}
&3 f^{(0,1)}(t,z) f(t,z) \phi ^{(0,1)}(t,z)+3 f^{(1,0)}(t,z) \phi ^{(1,0)}(t,z)+\frac{441 f(t,z)^2}{25 z^2}\\
&-\frac{8}{7} f(t,z)^2 \phi^{(0,1)}(t,z)^2-\frac{6 f(t,z)^2 \phi ^{(0,1)}(t,z)}{z}+4 f(t,z) \phi ^{(1,0)}(t,z)^2-6 f(t,z) \phi ^{(2,0)}(t,z)=0,\nonumber\\
&\nonumber\\
%Breathing Mode:
\label{98}
&3 z^2 f(t,z) f^{(0,1)}(t,z) \phi ^{(0,1)}(t,z)+3 z^2 f^{(1,0)}(t,z) \phi ^{(1,0)}(t,z)+\frac{7}{4} z^2 f^{(0,1)}(t,z)^2\\
&-\frac{7}{2} z^2 f(t,z)f^{(0,2)}(t,z)+\frac{7}{2} z f(t,z) f^{(0,1)}(t,z)-4 z^2 f(t,z)^2 \phi ^{(0,1)}(t,z)^2\nonumber\\
&+4 z^2 f(t,z) \phi ^{(1,0)}(t,z)^2+6 z^2 f(t,z)^2 \phi^{(0,2)}(t,z)-6 z^2 f(t,z) \phi ^{(2,0)}(t,z)+\frac{756}{25} f(t,z)^2=0.\nonumber
\end{align}
\newline
Let us assume the following leading behavior,
\begin{equation}
f_0(t,z)\sim f_0^{(\Delta_f)}(t)\ z^{\Delta_f}, \qquad \phi_0(t,z)\sim \frac{21}{10}\log z+\phi_0^{(\Delta_\phi)}(t)\ z^{\Delta_\phi}.
\end{equation}  
Plugging this ansatz into \eqref{94}-\eqref{98} and solving at leading order in the limit $z\rightarrow 0$, we get 
\begin{equation}
\Delta\equiv\Delta_f=\Delta_\phi=0 \quad \text{or} \quad \frac{14}{5}.
\end{equation}
Moreover $f_0^{(\Delta)}(t)$ and $\phi_0^{(\Delta)}(t)$ are left undetermined for both values of $\Delta$. Satisfying \eqref{94}-\eqref{98} at all subsequent orders in $z$ requires subleading terms in the background field expansions and imposes constraints among them. Quite importantly, one finds constraints (see \eqref{99}-\eqref{100}) on $f_0^{(\frac{14}{5})}(t)$ and $\phi_0^{(\frac{14}{5})}(t)$ coming from such subleading terms in the equations of motion. The resulting asymptotic expansion is
\begin{align}
\label{83}
f_0(t,z)=\,&z^0\left(f_0^{(0)}(t)+z^2 f_0^{(2)}(t)+z^4f_0^{(4)}(t)+\ldots\right)\\
&+z^{14/5}\left(f_0^{(\frac{14}{5})}(t)+z^2f_0^{(\frac{24}{5})}(t)+\ldots\right)\nonumber\\
&+z^{28/5}\left(f_0^{(\frac{28}{5})}(t)+z^2f_0^{(\frac{38}{5})}(t)+\ldots\right)\nonumber\\
&+z^{42/5}\left(f_0^{(\frac{42}{5})}(t)+z^2f_0^{(\frac{52}{5})}(t)+\ldots\right)\nonumber\\
&+z^{56/5}\left(f_0^{(\frac{56}{5})}(t)+z^2f_0^{(\frac{66}{5})}(t)+\ldots\right),\nonumber\\
%& \nonumber \\
\label{84}
\phi_0(t,z)=\,&\frac{21}{10} \log z\\
&+z^0\left(\phi_0^{(0)}(t)+z^2 \phi_0^{(2)}(t)\ldots\right)\nonumber\\
&+z^{14/5}\left(\phi_0^{(\frac{14}{5})}(t)+z^2\phi_0^{(\frac{24}{5})}(t)+\ldots\right)\nonumber\\
&+z^{28/5}\left(\phi_0^{(\frac{28}{5})}(t)+z^2\phi_0^{(\frac{38}{5})}(t)+\ldots\right)\nonumber\\
&+z^{42/5}\left(\phi_0^{(\frac{42}{5})}(t)+z^2\phi_0^{(\frac{52}{5})}(t)+\ldots\right)\nonumber\\
&+z^{56/5}\left(\phi_0^{(\frac{56}{5})}(t)+z^2\phi_0^{(\frac{66}{5})}(t)+\ldots\right).\nonumber
\end{align}
Let us comment on the structure of these expansions. Except for the logarithmic term in the dilaton expansion that can easily be recognized from \eqref{52}, there are five towers of terms, each of which has an internal spacing of two in powers of $z$. Such a spacing is very familiar from AdS-CFT and is referred to as Fefferman-Graham expansion \cite{Skenderis:2002wp}. It follows from the equations of motion \eqref{94}-\eqref{98} being second order differential in $z$. The first tower starts at order $z^0$ with some undetermined coefficients $f_0^{(0)}, \phi_0^{(0)}$ called \textit{sources} that need to be specified as Dirichlet conditions. Then the second tower starts at order $z^{14/5}$ with coefficients that are related to the expectation values of the operators dual to the considered bulk fields. In contrast to familiar examples of AdS-CFT dualities, their dependence on sources is local:
\begin{align}
\label{99}
f_0^{(\frac{14}{5})}(t)&=-\frac{9}{14}M_0\ f_0^{(0)}(t)\ e^{\frac{4}{3}\phi_0^{(0)}(t)},\\
\label{100}
\phi_0^{(\frac{14}{5})}(t)&=-\frac{3}{8}M_0\ e^{\frac{4}{3}\phi_0^{(0)}(t)},
\end{align}
with $M_0$ an undetermined constant that can be interpreted as a mass parameter characterizing the most general solution of the gravity-dilaton system \eqref{70}-\eqref{71} derived in Appendix~\ref{App:Background Solution}. The reason for this complete locality is that the dilaton and metric are non-propagating fields in this two-dimensional system. Finally the remaining towers are generated by the non-linearities of the equations of motion.\footnote{There are up to quartic terms in \eqref{94}-\eqref{98}, one can therefore get towers starting at $z^{14n/5}$ with $n=1,2,3,4$.} A similar structure has been displayed for a somewhat simpler system in \cite{vanRees:2011fr}. One should note that all coefficients depend locally on the sources $f_0^{(0)}, \phi_0^{(0)}$, although we will refrain from giving the precise relations. These can be easily recovered by inserting \eqref{83}-\eqref{84} in the equations of motion and solving order by order in $z$.

The asymptotics of fluctuations are similarly obtained. From the linearized equations of motion, a decoupled one can be derived for the breathing mode,
\begin{align}
\label{101}
&-\frac{1}{2} z^2 b_1^{(0,1)}(t,z) f_0^{(0,1)}(t,z)-\frac{b_1^{(1,0)}(t,z) f_0^{(1,0)}(t,z)}{2f_0(t,z)}+\frac{6}{7} z^2 b_1^{(0,1)}(t,z) f_0(t,z) \phi_0^{(0,1)}(t,z)\\
&-z^2b_1^{(0,2)}(t,z) f_0(t,z)-z b_1^{(0,1)}(t,z) f_0(t,z)-\frac{6}{7} b_1^{(1,0)}(t,z) \phi_0^{(1,0)}(t,z)+b_1^{(2,0)}(t,z)\nonumber\\
&+\frac{392}{25} b_1(t,z) f_0(t,z)=0.\nonumber
\end{align}
Performing a similar analysis as for the background fields, we find the following asymptotic expansion:
\begin{align}
\label{85}
b_1(t,z)=\,&z^{-14/5}\left(b_1^{(-\frac{14}{5})}(t)+z^2 b_1^{(-\frac{4}{5})}(t)+z^4 b_1^{(\frac{6}{5})}(t)+\ldots\right)\\
&+z^{0}\left(b_1^{(0)}(t)+z^2b_1^{(2)}(t)+\ldots\right)\nonumber\\
&+z^{14/5}\left(b_1^{(\frac{14}{5})}(t)+z^2b_1^{(\frac{24}{5})}(t)+\ldots\right)\nonumber\\
&+z^{28/5}\left(b_1^{(\frac{28}{5})}(t)+z^2b_1^{(\frac{38}{5})}(t)+\ldots\right)\nonumber\\
&+z^{42/5}\left(b_1^{(\frac{42}{5})}(t)+z^2b_1^{(\frac{52}{5})}(t)+\ldots\right).\nonumber
\end{align}
Again there are five towers of equally spaced terms in this expansion. The first one starts at order $z^{-14/5}$ with a \textit{source} that needs to be specified as Dirichlet boundary value. The fourth tower starts at order $z^{28/5}$ with a coefficient left undetermined by the asymptotic analysis. (In Euclidean signature, it is also a functional of the sources $\{f_0^{(0)},\phi_0^{(0)},b_1^{(-\frac{14}{5})}\}$, but the dependence is nonlocal and cannot be determined from a boundary asymptotic analysis only; specific solutions need to be known deeper in the bulk in order to fix its dependence in sources, which is in general not local or analytic. In Lorentzian signature, it also depends on the choice of state, or initial conditions, and encodes information on a \textit{propagating degree of freedom}.) The remaining towers are generated by the non-linearities of the equation of motion \eqref{101} with respect to the background fields. A particular point to note is that the coefficient $b_1^{(0)}$ is also undetermined but needs to be set to zero, as otherwise it would change the breathing mode background value in \eqref{76}.

Still at linear order in fluctuations, the dilaton and metric being non-propagating fields are constrained by the remaining equations of motions. In particular they enjoy the same power expansion,
\begin{align}
\label{90}
f_1(t,z)&=\sum\limits_{j} z^j f_1^{(j)}(t),\\
\label{91}
\phi_1(t,z)&=\sum\limits_{j} z^j \phi_1^{(j)}(t),
\end{align}
with running index given by $j\in\{-\frac{14}{5},\ldots,2,\ldots,\frac{14}{5},\ldots,\frac{28}{5},\ldots,\frac{42}{5},\ldots\}$. One can again check that all coefficients in \eqref{85}-\eqref{91} can be expressed as local functionals of the sources $\{f_0^{(0)},\phi_0^{(0)},b_1^{(-\frac{14}{5})}\}$ and the remaining undetermined coefficient $b_1^{(\frac{28}{5})}$. \footnote{Local dependences on $b_1^{(\frac{28}{5})}$ arise at order $j=\frac{28}{5}$ and further in the above series expansions.}

A striking feature of expansions \eqref{85}-\eqref{91} is that fluctuations seem to dominate over background fields in the limit $z \rightarrow 0$, and deform the geometry from being asymptotically $AdS_2$. This is the reason why we cannot consider a dynamical breathing mode non-perturbatively, and is mirrored by the fact that its dual matrix theory operator $T^{--}$ is an irrelevant one.

%%%%%%%%%%%%%%%%%%%%%%%%%%%%%%%%%%%%%%%%%%%%%%%%%%%%%%%%%%
\section{On-Shell Evaluation of the Action} \label{App:Onshell}
Considering fluctuations of the form
\begin{align}
\delta g_{\mu\nu}&=\epsilon\ h_{\mu\nu},\\
\phi&=\phi_0+\epsilon\ \phi_1,\\
b&=b_0+\epsilon\ b_1,
\end{align}
with constant $b_0=\frac{25}{4}$, we give the gauge invariant expression of the action \eqref{61} when evaluated on-shell, up to quadratic order in the expansion parameter $\epsilon$:
\newpage
\begin{align}
S_{on-shell}&=\frac{2\pi^{9/2}}{\Gamma \left(\frac{9}{2}\right)}\int d^{1}x\ \sqrt{-\gamma_0}\ e^{-\frac{6}{7}\phi_0} \hat{n}_{\mu}\left[F^{\mu}_{(0)}\left[g,\phi,b\right]+\epsilon F^{\mu}_{(1)}\left[g,\phi,b\right]+\epsilon^2F^{\mu}_{(2)}\left[g,\phi,b\right]\right]+\mathcal{O}(\epsilon^3)\nonumber\\
\label{11}
&+\frac{4\pi^{9/2}}{\Gamma \left(\frac{9}{2}\right)}\int d^{1}x\ \sqrt{-\gamma}\  e^{-\frac{6}{7}\phi} b^4 K,
\end{align}
with
\begin{align*}
F^{\mu}_{(0)}\left[g,\phi,b\right]=-\frac{16}{21}b_0^4\nabla^{\mu}\phi_0,
\end{align*}
\begin{align*}
F^{\mu}_{(1)}\left[g,\phi,b\right]=b_0^4\left[\frac{32}{49} \phi_1 \nabla^\mu \phi_0-\nabla^{\mu}h_\nu^\nu+\nabla^\nu h^{\mu}_{\nu}+\frac{6}{7} h^{\mu}_{\nu}\nabla^{\nu}\phi_0-\frac{6}{7} h_\nu^\nu \nabla^\mu \phi_0\right]-b_0^3\left[\frac{48}{7} b_1 \nabla^\mu \phi_0\right],
\end{align*}
\begin{align*}
F^{\mu}_{(2)}\left[g,\phi,b\right]&= b_0^4 \bigl[-\frac{96}{343} \phi_1^2 \nabla^{\mu } \phi_0 + \frac{16}{49} \phi_1 \nabla^{\mu }\phi_1 \bigr]\\
&+b_0^4 \Bigl[ \frac{3}{7} \phi_1 \nabla^{\mu }h_{\nu }^{\nu } + \frac{26}{49}h_{\nu }^{\nu } \phi_1 \nabla^{\mu }\phi_0 - \frac{3}{7} h_{\nu }^{\nu } \nabla^{\mu }\phi_1 - \frac{3}{7} \phi_1 \nabla_{\nu }h^{\mu \nu} -  \frac{34}{49} h_{\nu }^{\mu } \phi_1 \nabla^{\nu }\phi_0 \\
&\hspace{1cm}+ \frac{3}{7} h_{\nu }^{\mu } \nabla^{\nu }\phi_1 \Bigr]\\
&+b_0^4 \Bigl[- \frac{1}{4} h_{\sigma} ^{\sigma}  \nabla^{\mu }h_{\nu }^{\nu } +\frac{3}{4} h^{\nu \sigma}  \nabla^{\mu }h{}_{\nu \sigma}  + \frac{3}{7} h{}_{\nu \sigma}  h^{\nu \sigma}  \nabla^{\mu }\phi_0  - \frac{3}{14} h_{\nu }^{\nu } h_{\sigma} ^{\sigma}  \nabla^{\mu }\phi_0 + \frac{1}{4} h_{\sigma} ^{\sigma}  \nabla_{\nu }h^{\mu \nu} \\
&\hspace{1cm}+\frac{1}{2}  h^{\mu \nu} \nabla_{\sigma}h^{\sigma}{}_{\nu} -  \frac{1}{2} h^{\nu \sigma} \nabla_{\nu }h^{\mu }{}_{\sigma} -  \frac{3}{2} h^{\mu \sigma}  \nabla_{\nu }h^{\nu }{}_{\sigma}  + \frac{3}{4} h^{\mu \nu}  \nabla_{\nu }h_{\sigma} ^{\sigma}  -\frac{6}{7}\ h^{\mu }{}_{\sigma}  h_{\nu }^{\sigma}  \nabla^{\nu }\phi_0\\
&\hspace{1cm} + \frac{9}{14} h_{\nu }^{\mu } h_{\sigma} ^{\sigma}  \nabla^{\nu }\phi_0 \Bigr]\\
&+b_0^3 \bigl[-\frac{24}{7} \phi_1 \nabla^{\mu }b_1 + \frac{208}{49} b_1 \phi_1 \nabla^{\mu }\phi_0 -\frac{24}{7} b_1 \nabla^{\mu }\phi_1\bigr]\\
&+ b_0^3 \bigl[2 h_{\nu }^{\nu } \nabla^{\mu }b_1 - 2 b_1 \nabla^{\mu }h_{\nu }^{\nu } -\frac{24}{7} b_1 h_{\nu }^{\nu } \nabla^{\mu }\phi_0 + 2 b_1 \nabla_{\nu }h^{\mu \nu}  - 2 h_{\nu }^{\mu } \nabla^{\nu }b_1 +\frac{36}{7} b_1 h_{\nu }^{\mu } \nabla^{\nu }\phi_0\bigr]\\
&+b_0^2 \bigl[14 b_1 \nabla^{\mu }b_1 -\frac{72}{7} b_1^2 \nabla^{\mu }\phi_0 \bigr]. 
\end{align*}

We also note that the extrinsic curvature $K$ and the outward-pointing normal vector $\hat{n}^{\mu}$ have the following expressions in Fefferman-Graham gauge \eqref{44}:
\begin{align}
\hat{n}^{\mu}&=-z\delta_{z}^{\mu},\\
K&=-\frac{z}{2}f(t,z)^{-1}f'(t,z).
\end{align}

%%%%%%%%%%%%%%%%%%%%%%%%%%%%%%%%%%%%%%%%%%%%%%%%%%%%%%%%%%
\section{One-point Functions} \label{App:One-point}
In this Appendix, we compute the one-point functions of the operators dual to the considered bulk fields $g_{\mu,\nu},\ \phi$ and $b_1$, and explain why the encoded information about the matrix theory state is somewhat trivial in the sense that it is non-dynamical. As mentioned in the Introduction, sources for the irrelevant operator $T^{--}$ can be used in the generating functional of the matrix theory for computing $n$-point correlators, but have to be set to zero afterwards. From the holographic point of view, the breathing mode source $b_1^{source}\equiv b_1^{(-\frac{14}{5})}$ plays this role; see Appendix \ref{App:Asymptotics}. Using the expression for the renormalized on-shell action \eqref{27} we derive the various one-point functions,
\begin{align}
\label{113}
\langle \mathcal{O}_{\phi}(t) \rangle&= \frac{1}{\sqrt{f_0^{(0)}(t)}} \frac{\delta S^{ren}}{\delta \phi_0^{(0)}(t)}\Bigg|_{b_1^{source}=0}=\frac{\pi^{9/2}}{\Gamma \left(\frac{9}{2}\right)}\ \frac{5^9}{2^6\ .\ 3\ .\ 7}\ M_0\ e^{\frac{10}{21}\phi_0^{(0)}(t)},\\
\label{114}
\langle T^{tt}(t) \rangle&= \frac{2}{\sqrt{f_0^{(0)}(t)}} \frac{\delta S^{ren}}{\delta \left(-f_0^{(0)}(t)\right)}\Bigg|_{b_1^{source}=0}=-\frac{\pi^{9/2}}{\Gamma \left(\frac{9}{2}\right)}\ \frac{5^8}{2^7}\ M_0\ \frac{e^{\frac{10}{21}\phi_0^{(0)}(t)}}{f_0^{(0)}(t)},\\
\label{115}
\langle \mathcal{O}(t) \rangle&= \frac{1}{\sqrt{f_0^{(0)}(t)}} \frac{\delta S^{ren}}{\delta\ \epsilon e^{-\frac{6}{7} \phi_0^{(0)}}b_1^{source}}\Bigg|_{b_1^{source}=0}=\frac{\pi^{9/2}}{\Gamma \left(\frac{9}{2}\right)}\ \frac{5^{13}\ .\ 13}{2^{10} \ .\ 7^2}\ M_0^2\ e^{\frac{8}{3}\phi_0^{(0)}(t)}.
\end{align}

One can check that the following trace and diffeomorphism Ward identities are satisfied \cite{Kanitscheider:2008kd}:
\begin{align}
\frac{21}{10}\langle \mathcal{O}_{\phi}(t) \rangle-\langle T^{t}_t(t) \rangle=0,\\
\nabla^t\langle T_{tt}(t)\rangle-\partial_t \phi_0^{(0)}(t)\ \langle \mathcal{O}_{\phi}(t) \rangle=0.
\end{align}
This actually is a simple consequence of \eqref{99}-\eqref{100}.

Let us briefly comment on \eqref{113}-\eqref{115}. As can be seen, all these one-point functions simply follow the dilaton source profile $\phi_0^{(0)}(t)$. Looking at a quenched solution of the type \eqref{72}-\eqref{73} that is dual to sudden energy injection in the dual matrix theory groundstate, we see that one-point functions simply jump between zero and a constant set by the mass parameter $M_0$ (related the to final temperature through equation \eqref{81}). Expectation values instantaneously take their thermal value and no dynamical thermalization can be seen from these observables. In order to see dynamical thermalization in matrix theory, we therefore need to look at higher-point correlators like the one we study in Section \ref{section:Linear Response}.

%%%%%%%%%%%%%%%%%%%%%%%%%%%%%%%%%%%%%%%%%%%%%%%%%%%%%%%%%%
\section{Boundary-to-Bulk Propagators in Vacuum $AdS_2$} \label{App:Propagators}
We review the derivation of breathing mode boundary-to-bulk propagators in pure $AdS_2$. For convenience, we start in Poincar\'e coordinates and then give the expression in Eddington-Finkelstein gauge.

In the Euclidean version of vacuum $AdS_2$ described by the metric
\begin{align}
ds^2=\frac{1}{z^2}\left(d\tau^2+dz^2\right),
\end{align}
the unique breathing mode boundary-to-bulk propagator is given by
\begin{equation}
G_E^{AdS}(\tau,z)=\frac{\Gamma\left(\frac{47}{10}\right)}{\sqrt{\pi}\ \Gamma\left(\frac{21}{5}\right)}\frac{z^{28/5}}{(z^2+\tau^2)^{47/10}}.
\end{equation}
Indeed, one can show that it satisfies the equation of motion \eqref{17} and is normalized to a delta function source at the boundary, 
\begin{equation}
\lim\limits_{z\rightarrow 0}z^{14/5}\int_{-\infty}^{\infty}d\tau\ G_E^{AdS}(\tau,z)=1.
\end{equation}
\newline
It is well-known that this unique Euclidean boundary-to-bulk propagator corresponds to the Feynman time-ordered one after Wick rotation. For this we perform the analytic continuation $\tau=e^{i\theta}t$ from $\theta=0$ to $\theta=\frac{\pi}{2}$ (see \cite{Balasubramanian:2012tu} for a similar treatment) and obtain
\begin{align}
\label{12}
iG_F^{AdS}(t,z)&=-\frac{\Gamma\left(\frac{47}{10}\right)}{\sqrt{\pi}\ \Gamma\left(\frac{21}{5}\right)}\frac{z^{28/5}}{(z^2+e^{2i\theta}t^2)^{47/10}}\\
&=-\frac{\Gamma\left(\frac{47}{10}\right)}{\sqrt{\pi}\ \Gamma\left(\frac{21}{5}\right)}\ z^{28/5}\left(\frac{\theta(z^2-t^2)}{(z^2+e^{2i\theta}t^2)^{47/10}}+e^{-i \frac{47}{5}\theta}\frac{\theta(t^2-z^2)}{(t^2+e^{-2i\theta}z^2)^{47/10}}\right)\\
\label{26}
&=-\frac{\Gamma\left(\frac{47}{10}\right)}{\sqrt{\pi}\ \Gamma\left(\frac{21}{5}\right)}\ z^{28/5}\left(\frac{\theta(z^2-t^2)}{(z^2-t^2)^{47/10}}+e^{-i \frac{47}{10}\pi}\frac{\theta(t^2-z^2)}{(t^2-z^2)^{47/10}}\right),
\end{align}
where the overall minus sign is needed in order to satisfy
\begin{equation}
\label{13}
\lim\limits_{z\rightarrow 0}z^{14/5}\int_{-\infty}^{\infty}dt\ G_F^{AdS}(t,z)=1.
\end{equation}
\newline
We then write the retarded propagator as
\begin{align}
\label{38}
G_R^{AdS}(t,z)=2\ \theta(t)\ \text{Re}\left[ G_F^{AdS}(t,z) \right]=\frac{2\Gamma\left(\frac{47}{10}\right)}{\sqrt{\pi}\ \Gamma\left(\frac{21}{5}\right)}\ \sin\left(\frac{47\pi}{10}\right)\theta\left(t-z\right) \frac{z^{28/5}}{(t^2-z^2)^{47/10}},
\end{align}
which is correctly normalized at the boundary. Note that the retarded propagator vanishes outside the future lightcone as expected. By performing the variable change $v\equiv t-z$, we find the expression of the Feynman and retarded boundary-to-bulk propagators in ingoing Eddington-Finkelstein coordinates,
\begin{align}
ds^2=-\frac{1}{z^2}\left(2\ dvdz+dv^2\right),
\end{align} 
\begin{align}
G_{F}^{AdS}(v,z)&=\frac{i\Gamma\left(\frac{47}{10}\right)}{\sqrt{\pi}\ \Gamma\left(\frac{21}{5}\right)}\frac{e^{-i\frac{47}{10}\pi}\ z^{28/5}}{(v(v+2z))^{47/10}},\\
G_R^{AdS}(v,z)&=\frac{2\Gamma\left(\frac{47}{10}\right)}{\sqrt{\pi}\ \Gamma\left(\frac{21}{5}\right)}\ \sin\left(\frac{47\pi}{10}\right)\theta\left(v\right) \frac{z^{28/5}}{(v(v+2z))^{47/10}}.
\end{align}

\section{Lowest Quasinormal Modes}  \label{App:LowestQNM}
In this appendix we numerically compute the first and second quasinormal frequencies of breathing mode fluctuations around static black holes at temperature $T_H$ and show that the lowest one matches the single complex frequency dominating the late-time behavior of the leading coefficient \eqref{79} of the retarded boundary-to-bulk propagator  when considered on the collapsing thin shell background solution \eqref{67}-\eqref{67a}. We also derive the universality of frequency-to-temperature ratios that the results  \eqref{77}-\eqref{78} were indicating, using an argument given in \cite{Horowitz:1999jd}. For reviews of quasinormal modes, one can consult \cite{Kokkotas:1999bd,Konoplya:2011qq,Berti:2009kk}.

We therefore need to determine the lowest QNM frequencies of linear breathing mode perturbations around a static black hole solution. The discussion will closely follow that of \cite{Horowitz:1999jd} since the asymptotic geometry in both cases is that of AdS.

The static black hole background considered here is the near-horizon limit (or decoupling limit) of the black D-particle background found in \cite{Horowitz:1991cd}. Its expression in the dual frame is given by\footnote{Note that the coordinate $r$ here is different from that defined in equation \eqref{64}.\\}
\begin{align}
ds^2_{dual}&=-N(r)dt^2+N(r)^{-1}dr^2,\\
N(r)&=r^2\left[1-\left(\frac{r_h}{r}\right)^{14/5}\right],
\end{align}
together with the associated background dilaton
\begin{equation}
\phi(r)=-\frac{21}{10}\log r.
\end{equation}
In these coordinates, the AdS boundary is located at $r \rightarrow \infty$. The horizon radius of such a black hole is related to its mass through $r_h=M^{5/14}$, while its Hawking-Unruh temperature is given by \eqref{81}. In this coordinate system the decoupled equation of motion satisfied by the linearized breathing mode perturbation is 
\begin{equation}
\label{33}
\ddot{b_1}(t,r)-N(r)^2\ b_1''(t,r)+\left(\frac{6}{7}N(r)^2\phi'(r)-N(r)N'(r)\right)b_1'(t,r)+\frac{392}{25} N(r)\ b_1(t,r)=0.
\end{equation}

\subsection{Universality of Frequency-to-Temperature Ratio} \label{subsection:Universality}
Let us have a closer look at the breathing mode equation of motion \eqref{33}. By performing the coordinate rescaling $t=a\hat{t}, r=\hat{r}/a$, it becomes 
\begin{equation}
\label{34}
\ddot{b_1}(\hat{t},\hat{r})-\tilde{N}(\hat{r})^2\ b_1''(\hat{t},\hat{r})+\left(\frac{6}{7}\tilde{N}(\hat{r})^2\phi'(\hat{r})-\tilde{N}(\hat{r})\tilde{N}'(\hat{r})\right)b_1'(\hat{t},\hat{r})+\frac{392}{25} \tilde{N}(\hat{r})\ b_1(\hat{t},\hat{r})=0,
\end{equation}
where dots and primes refer now to derivatives with respect to $\hat{t}$ and $\hat{r}$, and where we have defined the function
\begin{equation}
\tilde{N}(\hat{r})\equiv \hat{r}^2\left[1-\left(\frac{ar_h}{\hat{r}}\right)^{14/5}\right].
\end{equation}

It is obvious that \eqref{34} has the exact same form as \eqref{33} up to the replacement $r_h\rightarrow \hat{r}_h\equiv ar_h$. This has the important consequence that the quasinormal frequency spectrum scales with the horizon radius $r_h$, or equivalently with the black hole temperature, which is the only scale in the problem. To see this, it suffices for example to consider the time-dependent part $e^{-i \omega\left[r_h\right] t}=e^{-i\omega\left[\hat{r}_h/a\right]\ a\hat{t}}$ of such a quasinormal mode, which has to be equal to $e^{-i\omega\left[\hat{r}_h\right]\hat{t}}$ from the form invariance of the equation of motion, and therefore of its solutions. We conclude that the frequency must be linear in $r_h$, or equivalently
\begin{equation}
\frac{\omega}{T_H}=constant,
\end{equation}
where the constant is different for each of the quasinormal modes forming a discrete spectrum.

\subsection{Frobenius Expansion Near the Horizon}
Using the following ansatz for the breathing mode perturbation:
\begin{equation}
b_1(t,r)=r^{-9/10}\chi(r) e^{-i \omega t},
\end{equation}
and defining the tortoise coordinate $r_*$ through $\frac{dr}{dr_*}=N(r)$, one arrives at a Schrödinger-like equation 
\begin{align}
\frac{d^2}{dr_*^2}\chi+(\omega^2-V)\chi=0,
\end{align}
with effective potential given by
\begin{equation}
V(r)=N(r)\left[\frac{392}{25}-\frac{9}{100}\frac{N(r)}{r^2}+\frac{9}{10}\frac{N'(r)}{r}\right].
\end{equation}

In order to study numerically the behavior of ingoing modes at the horizon, we factorize the associated singular part in the solution:
\begin{equation}
\chi=e^{-i \omega r_*}\zeta,
\end{equation} 
and end up with the equation of motion
\begin{equation}
\label{31}
\left[s(z)\frac{d^2}{dz^2} +\frac{t(z)}{(z-h)}\frac{d}{dz}+\frac{u(z)}{(z-h)^2}\right]\zeta=0,
\end{equation}
\begin{align}
s(z)&\equiv -\frac{N z^4}{z-h},\\
t(z)&\equiv -\left[2Nz^3-N'z^2+2i\omega z^2\right],\\
u(z)&\equiv \frac{V}{N}(z-h),
\end{align}
where we are now using Fefferman-Graham coordinates with $z=1/r$ and where we define the inverse horizon radius $h\equiv1/r_h$. One can check that the functions $s(z)$, $t(z)$ and $u(z)$ are regular between the horizon ($z=h$) and spatial infinity ($z=0$), so that $\zeta$ admits a Frobenius series expansion close to the horizon:
\begin{equation}
\label{32}
\zeta(z)=(z-h)^\alpha \sum_{n=0}^{\infty} a_n(\omega,h) (z-h)^n,
\end{equation}
where $\alpha$ is a complex parameter which should have two possible values, corresponding to ingoing and outgoing solutions.

In order to determine $\alpha$ and the coefficients $\{a_n(\omega,h)\}_{n \in \mathbb{N}}$, we also expand the known functions $s(z), t(z)$ and $u(z)$ around $z=h$:
%\newpage
\begin{align}
s(z)&=\sum_{i=0}^{\infty} s_i(\omega,h) (z-h)^i,\\
t(z)&=\sum_{i=0}^{\infty} t_i(\omega,h) (z-h)^i,\\
\label{39}
u(z)&=\sum_{i=0}^{\infty} u_i(\omega,h) (z-h)^i.
\end{align}
Then, by plugging \eqref{32}-\eqref{39} in  \eqref{31} and looking at the lowest order in $(z-h)$, one gets the \textit{indicial equation}
\begin{equation}
s_0\ \alpha(\alpha-1)+t_0\ \alpha +u_0=0.
\end{equation}
By using the known coefficients $s_0=2\kappa h^2, t_0=2(\kappa-i \omega)h^2, u_0=0$ where $\kappa\equiv N'(r_h)/2$ is the surface gravity at the horizon, one gets two independent solutions for $\alpha$:
\begin{align}
\text{ingoing solution}: \qquad  \alpha&=0,\\
\text{outgoing solution}: \qquad  \alpha&=i\omega/\kappa.
\end{align}  
One can indeed identify those solutions as ingoing and outgoing ones by noting that close to the horizon \cite{Horowitz:1999jd}
\begin{equation}
r_*=\int dr\ N(r)^{-1}\simeq \frac{1}{N'(r_h)} \log (r-r_h)=\frac{\kappa}{2}\log \left(\frac{1}{z}-\frac{1}{h}\right),
\end{equation}
such that ingoing and outgoing modes indeed match the two solutions we have found:
\begin{align}
\text{ingoing solution}: \quad b_1(t,r_*)&\sim e^{-i\omega(t+r_*)},\\
\text{outgoing solution}: \quad b_1(t,r_*)&\sim e^{-i\omega(t-r_*)}=e^{-i\omega(t+r_*)}e^{2i\omega r_*}\nonumber\\
&\simeq e^{-i\omega(t+r_*)} (z-h)^{i\omega/\kappa}
\end{align}

\newpage
Since we are interested in quasinormal modes, we impose that nothing comes out of the horizon by choosing $\alpha=0$. With this choice, we can return to \eqref{31} and solve iteratively for the $a_n(\omega,h)$, leading to the recursive formula \cite{Horowitz:1999jd}
\begin{equation}
a_n=-\frac{1}{n(n-1)s_0+nt_0}\sum_{k=0}^{\infty}\left[k(k-1)s_{n-k}+kt_{n-k}+u_{n-k}\right]a_k.
\end{equation} 

\subsection{Numerical Shooting Method} \label{subsection:NumShoot}
We will determine numerically the lowest quasinormal complex frequency
\begin{equation}
\omega\equiv\omega_R - i \omega_I,
\end{equation}
with $\omega_R, \omega_I \in \mathbb{R}$. From the positivity of $V(r)$ outside the horizon radius, it can be proven that $\omega_I$ must be a strictly positive real number, such that only decaying solutions are allowed \cite{Horowitz:1999jd}.

A simple method is to evaluate the Frobenius expansion \eqref{32} of $\zeta(z)$ with $\alpha=0$, at a point close to the horizon where it is valid,\footnote{The radius of convergence of the Frobenius series is at least equal to $|h-p|$, where $p$ is the singular point of \eqref{31} which is the closest one to $h$ in the complex $z$-plane. In this case the closest singular point is $e^{i\pi/7}h$, such that the radius of convergence is at least $0.445\ h$. We have chosen $z_{ini}=0.75\ h$ as initial point for our numerical integration, which therefore lies in the convergence disk centered around $z=h$.} and then integrate \eqref{31} numerically  up to spatial infinity, for various values of $\omega$. Then quasinormal frequencies are those for which the solution is normalizable (does not diverge) at spatial infinity. Of course one has to truncate the Frobenius expansion at some order $N$ and check for the convergence of the results for increasing $N$. For the purpose of this work, it was sufficient to truncate the series at order $N=40$.

Using this numerical shooting method, we give results for the first and second lowest quasinormal frequencies:
\begin{align}
\text{First QNM:} \qquad &\frac{\omega_R}{T_H}=13.7,\\
&\frac{\omega_I}{T_H}=26.7,
\end{align}
\begin{align}
\text{Second QNM:} \qquad &\frac{\omega_R}{T_H}=18.6,\\
&\frac{\omega_I}{T_H}=36.6.
\end{align}
As one can see, the value of the first quasinormal frequency is indeed in good agreement with the single complex frequency \eqref{77}-\eqref{78} dominating the  leading coefficient  of the retarded boundary-to-bulk propagator, whose time evolution is computed in Section \ref{section:Propagators}.

%%%%%%%%%%%%%%%%%%%%%%%%%%%%%%%%%%%%%%%%%%%%%%%%%%%%%%%%%

\end{document}